# Determining the chemical composition of diamagnetic mixed solids via measurements of the magnetic susceptibility

Miao Miao Zhao,† Yang Yang,† Na Du, Yu Ying Zhu, Peng Ren,* Fei Yen*

*School of Science, Harbin Institute of Technology,*
*Shenzhen, University Town, Shenzhen, Guangdong 518055 P. R. China.*

†These authors contributed equally;
*Email: fyen@hit.edu.cn, renpeng@hit.edu.cn

**Abstract:** Mixed solid compounds are employed in a vast array of applications so an accurate determination of their chemical compositions is of crucial importance. All current characterization methods require specially-treated samples so the availability of a more practical method with similar accuracy should alleviate the quantification process. In this work, we show how the doping concentration δ (or isotope concentration) of a mixed solid compound in powdered form, where both parent compounds are diamagnetic, can be obtained from the measurement of the mass magnetization. We exploit the additive nature of the molar magnetic susceptibility $\chi_{Mol}$ and molar mass to construct two equations with the same two unknowns in the $\chi_{Mol}$ vs. δ space to simultaneously solve $\chi_{Mol}$ and δ of a mixed solid. Eight examples are provided to show the wide applicability of this method: $NH_{4(1-\delta)}D_{4\delta}Br$ (where D = $^{2}$H), $NH_4I_{1-\delta}Br_\delta$, $(NH_4H_2)_{1-\delta}(ND_4D_2)_\delta PO_4$, $C_{48}H_{22+6\delta}Br_{6(1-\delta)}O_{32}Zr_6$, $[creatine]_{1-\delta}[$D-glucose$]_\delta$, $[$L-glutamic acid$]_{1-\delta}[$L-leucine$]_\delta$, $[terephthalic\ acid]_{1-\delta}[trimesic\ acid]_\delta$ and $[$*p*-terphenyl$]_{1-\delta}[triphenylphosphine]_\delta$. Experimental errors of ±1.2% were obtained for δ from average sample masses of 16.6 mg in powdered form rendering the presented approach an attractive choice for characterizing the ratios of mixed solids.

## Introduction

Replacement of a fraction of the cations A and B or anions X of solid compounds of the types AX or ABX by a different element of the same valency A', B' or X', respectively, can dramatically alter the physical and chemical properties of the system.[1-5] The ions can also be replaced by the same element but with a different isotope.[6] During the synthesis process of a sample with such mixed concentrations, often the chemical composition of the final product is different than the molar ratios of the starting solution. For example, a solution comprised of deionized $H_2O$ with 85% of $NH_4I$ and 15% of KI powder yields crystals of the composition $NH_4I_{0.83}K_{0.17}$.[7] To determine the exact chemical composition, X-ray diffraction (XRD) measurements can be performed on the sample with unknown $NH_4^+$ and $K^+$ concentrations to obtain the average lattice constant(s) and matching it to a reference. The reference is usually a straight line connecting the lattice constants of the pure compounds $NH_4I$ and KI *i.e.*, Vegard's Law.[8] However, often the lattice constant(s) of even perfectly miscible compounds do not follow the additive rule.[9] As such, another type of measurement technique to determine the doping concentration of a system is required.

The replacement ions A', B' and X' can also be of the same element but different isotope.[10] Isotope dating of natural occurring compounds such as rocks, meteorites and ice is already a mature field of study.[11,12] In contrast, research on deliberate synthesis of compounds with mixed isotopes remains in its infancy.[6,10] Nevertheless, more and more applications with mixed isotopes are flourishing and accurate determination of their isotope concentrations is increasing in demand. For instance, precision deuteration of pharmaceutical drugs may favorably alter their pharmacokinetics and metabolic rates;[13] selective substitution of hydrogen by deuterium can result in different photoelectronic responses in conducting polymers;[14,15] and partial replacement of H by D, such as $NH_{3.36}D_{0.64}Br$, induces a ferroelectric phase.[16] The two most common methods to determine the isotope concentrations of a material are IRMS (isotope ratio mass spectroscopy) and NMR spectroscopy. IRMS requires thermal ionization of the samples while NMR requires 'digesting' of samples and expensive equipment. From such, an alternative method to determining the isotope concentration of a material that is non-destructive and/or more economically accessible is also desirable.

Crystal growth of many organic compounds is rather difficult. This is also true for the case of mixed metal-organic frameworks (MOFs) employed in a large variety of applications.[17,18] In the cases when mixed crystals can be grown, due to their structures being nearly identical to the parent compounds, X-ray diffraction patterns are often indifferentiable so quantifying the chemical compositions of mixed MOFs is challenging. Some ordinary organic compounds also pose the challenge in that they possess two stable polymorphs under ambient conditions. In most cases the two polymorphs have different structures that behave plastically such as creatine, glucose and a plethora of pharmaceutical drugs.[19,20] For these reasons, working with lattice constants via XRD measurements to determine chemical composition is not feasible in such situations. Moreover, determination of the chemical composition of mixed organic compounds via mass spectroscopy and X-ray fluorescence is challenging because, respectively, the parent compounds may yield similar fragmented ions and the fluorescence of lighter atoms ($Z < 11$) is rather weak. This calls for the need of an alternative method capable of determining the stoichiometric ratio of mixed organic compounds, preferably one that does not require samples to be in crystalline form or digestion of large amounts of samples.

All solid compounds exhibit diamagnetism and their overall diamagnetic signal (manifested in the magnetic susceptibility) is comprised of additive components known as Pascal's constants.[21] The additive components are based on the compounds' atomic, molecular and bonding compositions. Indeed, in binary compounds, at least experimentally, the molar magnetic susceptibility varies linearly with the concentration of the two pure compounds.[22,23] However, applying such a linear relationship to determine the concentration of a mixed compound has been deemed impossible since in order to obtain the molar magnetic susceptibility of the mixed sample, its molar mass is needed which is unknown to begin with. Nevertheless, determination of the concentration is still possible by also exploiting the linear relationship of the molar masses. The process is not straight forward and to our knowledge, no literature employing such a method to determining the doping (and isotope) concentrations of mixed compounds have previously been reported. A good portion of inorganic compounds and *most* organic compounds are diamagnetic so the presented method is expected to have wide applicability.

In this work, we first briefly review the topic of diamagnetism. Then, we describe the method of extracting the doping (or isotope) concentration $\delta$ of mixed solids via measurements of the magnetic susceptibility. Three examples are provided for inorganic compounds, one for each case where the lattice constant 1) follows the additivity rule; 2) only partially follows the additivity rule; and 3) does not follow the additivity rule: these three systems are, respectively, $NH_{4(1-\delta)}D_{4\delta}Br$, $NH_4I_{1-\delta}Br_\delta$ and $(NH_4H_2)_{1-\delta}(ND_4D_2)_\delta PO_4$. Around half of metal-organic frameworks are diamagnetic so the system $C_{48}H_{22+6\delta}Br_{6(1-\delta)}O_{32}Zr_6$ was selected as a fourth example. Furthermore, four examples are provided for organic compounds: [creatine]$_{1-\delta}$[D-glucose]$_\delta$, [L-glutamic

acid]$_{1-\delta}$[L-leucine]$_{\delta}$, [terephthalic acid]$_{1-\delta}$[trimesic acid]$_{\delta}$ and [*p*-terphenyl]$_{1-\delta}$[triphenylphosphine]$_{\delta}$. These were selected from commonly used reagents that can partially represent the large variety of organic compounds comprised of metal organic frameworks, acids, sugars and ligands. Laslty, we discuss how the error margins of δ were extracted. On average, δ was resolved to 83 parts from only working with 16.6 mg of powdered samples. Higher resolutions can be obtained by increasing sample masses since they scale linearly with each other.

## Brief review of diamagnetism

All solids exhibit diamagnetism because its paired electrons, while confined to their parent atoms or bonds, possess orbital degrees of freedom that allow them to respond to an applied magnetic field $H$. These electrons end up orbiting near their parent atom or bond in a fashion as to generate a magnetic field opposite to $H$ in the attempts of maintaining the local magnetic field unchanged, *i.e.*, Lenz's law. Some paired electrons reside deep within atomic shells and others constitute covalent bonds so their diamagnetic response is different which may lead to diamagnetic anisotropy[24] and exalted diamagnetism.[25] The contributions from the paired electrons of each type of atom and covalent bond to the molar magnetic susceptibility $\chi_{\mathrm{Mol}}$ are commonly known as Pascal's constants[26] in units of cm$^3$/mol. The addition of all of the Pascal's constants of a molecule provides a good approximation of the system's expected $\chi_{\mathrm{Mol}}$. Note that if there is a presence of unpaired electrons, then the system is paramagnetic at high temperatures and below a critical temperature ferromagnetic or antiferromagnetic; and their contributions to $\chi_{\mathrm{Mol}}$ far exceed that of the paired electrons. For the sake of simplicity, we will only discuss materials that do not have unpaired electrons.

## Method of obtaining δ from the diamagnetic susceptibility

Magnetometers usually measure the total magnetic moment $\boldsymbol{m}$ of a sample in units of emu. Here, $\boldsymbol{m} = \Sigma\ \mu$, is the summation of all of the individual magnetic moments $\mu$ within the sample. For diamagnetic systems, $\boldsymbol{m}$ is negative because the directions of all $\mu$ are in the opposite direction of $\boldsymbol{H}$. The larger the sample the larger $\boldsymbol{m}$ is, so more useful quantities are the mass, volume and molar magnetizations, $\boldsymbol{M}_{\mathrm{Mass}}$, $\boldsymbol{M}_{\mathrm{Vol}}$, $\boldsymbol{M}_{\mathrm{Mol}}$, in units of emu/g, emu/cm$^3$ and emu/mol, respectively. The magnetic susceptibility is treated to behave as $\chi = \boldsymbol{M}/\boldsymbol{H}$ for diamagnetic systems. Hence, even more useful quantities are the analogous mass, volume and molar susceptibilities derived from $\chi_{\mathrm{mass}} = \boldsymbol{M}_{\mathrm{Mass}}/\boldsymbol{H}$, $\chi_{\mathrm{Vol}} = \boldsymbol{M}_{\mathrm{Vol}}/\boldsymbol{H}$ and $\chi_{\mathrm{Mol}} = \boldsymbol{M}_{\mathrm{Mol}}/\boldsymbol{H}$, respectively, in units of emu/g-Oe, emu/cm$^3$-Oe and emu/mol-Oe. These are the true intrinsic values characterizing a diamagnetic system with $\chi_{Mol}$ being the most valuable one because it has the same units as those of Pascal's constants (emu/mol-Oe = $cm^3$/mol) and because only $\chi_{Mol}$ is additive, whereas $\chi_{mass}$ and $\chi_{Vol}$ are not. Given the linear relationship between $\chi$ and $\boldsymbol{H}$ (for diamagnets), $\chi_{Mol}$ is easily obtainable from the slope of a linear fit of the $\boldsymbol{M}_{Mol}$ vs. $\boldsymbol{H}$ plot. Experiments can be performed in moderate magnetic fields of −10 kOe to +10 kOe and at room temperature since $\chi$ is independent of temperature for diamagnetic systems. For the parent compounds, their $\boldsymbol{M}_{Mol} = \boldsymbol{m} \cdot [m_{Mol} / m]$ are easily derived because the masses of the samples $m$ and molar masses $m_{Mol}$ (in units of g/mol) are known. From such, $\chi_{Mol_A}$ and $\chi_{Mol_A'}$ of the parent compounds A and A' are also easily obtained. Since the electrons only interact with $\boldsymbol{H}$ and not with each other, it is presumed that all samples with mixed concentrations $A_{1-\delta}A'_{\delta}$ must lie within a straight Line 1 connecting the points (0, $\chi_{Mol_A}$) and (1, $\chi_{Mol_A'}$) in the $\chi_{Mol}$ vs. $\delta$ space (Figure 1). Line 1 has the following relationship:

$$\chi_{Mol} = (\chi_{Mol_A'} - \chi_{Mol_A}) \cdot \delta + \chi_{Mol_A} \qquad (1)$$

The problem arises when we attempt to determine $\boldsymbol{M}_{Mol}$ (and therefore $\chi_{Mol}$) of a mixed sample with unknown $\delta$. At first glance, we would presume that $\delta$ can be obtained from Line 1 (in Fig. 1) by simply measuring $\chi_{Mol}$ of the mixed sample. However, this is not possible because to obtain $\chi_{Mol}$ we need to know the molar mass of the mixed sample, which depends on $\delta$, and $\delta$ is unknown to begin with. To circumvent this dilemma, note that $\boldsymbol{M}_{Mass}$ is measurable so $\chi_{Mass}$ can be determined. The relationship between $\chi_{Mol}$ and $\chi_{Mass}$ is $\chi_{Mol} = \chi_{Mass} \cdot m_{Mol}$. While the exact value of $m_{Mol}(\delta)$ is unknown for the mixed sample, we can still be certain that $m_{Mol}(\delta)$ is a superposition of the parent compounds, where $m_{Mol}(\delta) = m_{Mol_A} \cdot (1 - \delta) + m_{Mol_A'} \cdot \delta$, which upon substitution yields a second condition, Line 2:

$$\chi_{Mol} = \chi_{Mass} \cdot [m_{Mol_A} \cdot (1 - \delta) + m_{Mol_A'} \cdot \delta] \qquad (2),$$

The physical meaning of Equation (2) is the following: it represents all of the possible $\chi_{Mol}$ values of a mixed sample belonging to the $A_{1-\delta}A'_{\delta}$ series with known $\chi_{Mass}$, $m_{Mol_A}$ and $m_{Mol_A'}$ but unknown $\delta$, in the span between 0 and 1 (dotted line shown in Fig. 2). While any point not resting on Line 2 is not a solution, there exists an infinite number of combinations in Equation (2). However, upon imposing the restriction of Line 1, meaning that $\chi_{Mol}$ is also restricted to have to lie on Line 1, then only one unique solution can simultaneously satisfy both conditions: the point of intersection of Line 1 and Line 2.

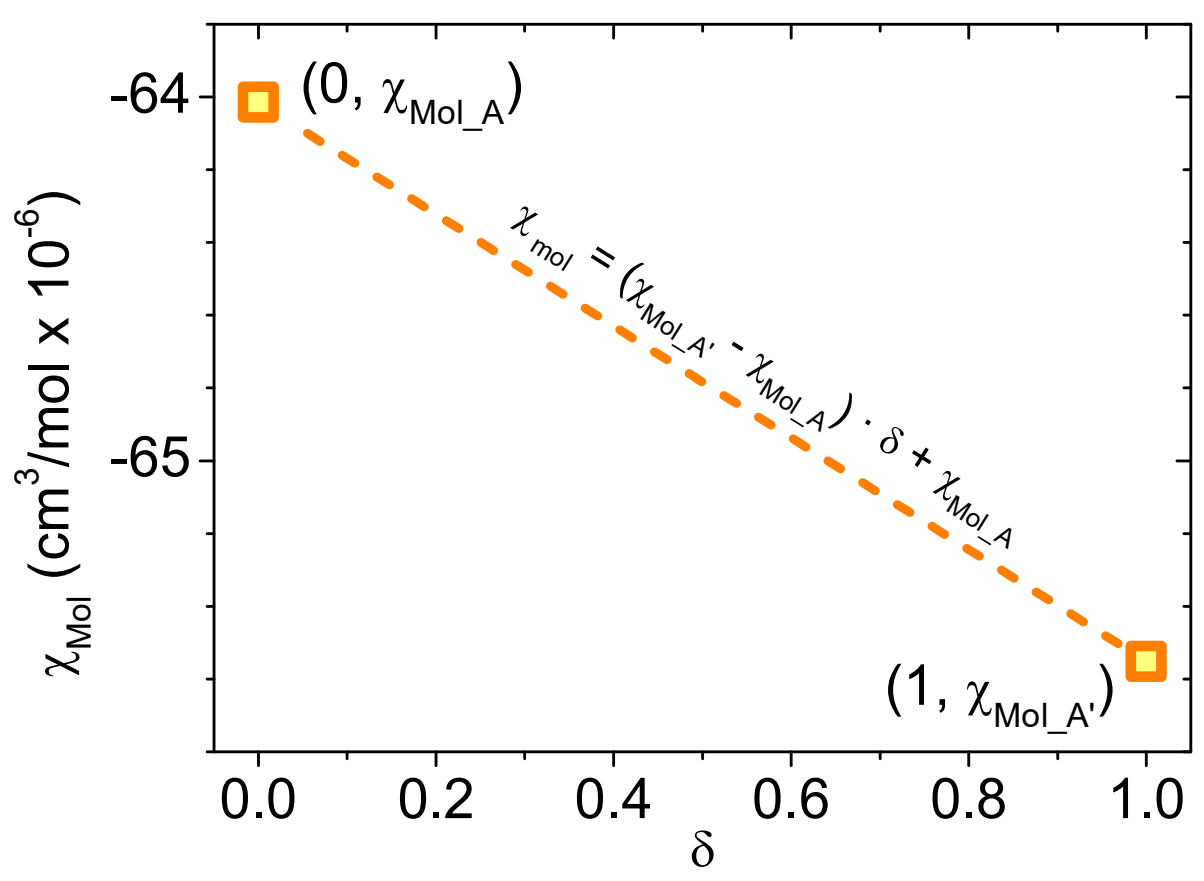

**Fig. 1.** Relationship between doping concentration δ and molar magnetic susceptibility $\chi_{Mol}$ of the series $A_{1-\delta}A'_{\delta}$. $\chi_{Mol}$ of the two pure compounds A and A' are determinable because δ is known. Dashed line connecting (0, $\chi_{Mol_A}$) and (1, $\chi_{Mol_A'}$) is Eq. 1.

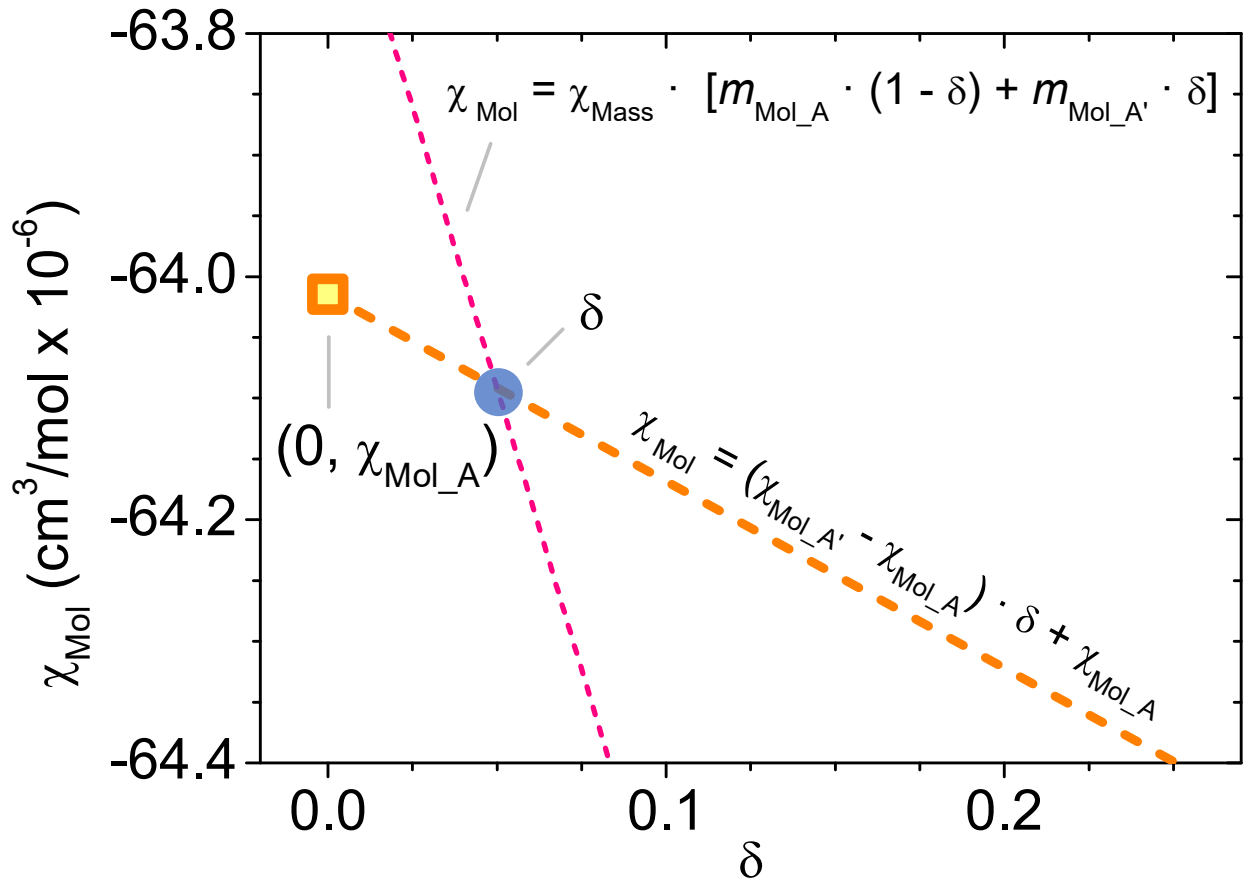

**Fig. 2.** For mixed samples, only employing Eq. 1 to obtain δ is not feasible because to determine $\chi_{Mol}$, δ also needs to be known. A second line (dotted) in the form of Eq. 2 is introduced so now there are two equations with the same two unknown variables: δ and $\chi_{Mol}$.

## Results and discussion

### Example 1: $NH_{4(1-\delta)}D_{\delta}Br$

The molar magnetic susceptibilities obtained for $NH_4Br$ and $ND_4Br$ (98% deuterated) were $\chi_{Mol\text{-}NH4Br} = -49.036(6)\times10^{-6}$ cm$^3$/mol and $\chi_{Mol\text{-}ND4Br_98\%} = -51.674(14)\times10^{-6}$ cm$^3$/mol, respectively. From such, the corresponding points (0, $\chi_{Mol\text{-}NH4Br}$) and (0.98, $\chi_{Mol\text{-}ND4Br}$) are plotted in Fig. 3. The dashed line connecting these two points is Line 1 for the $NH_{4(1-\delta)}D_{4\delta}Br$ series: $\chi_{Mol} = [-2.638(20) \cdot \delta - 49.036(6)]\times10^{-6}$ cm$^3$/mol. This implies that all samples in the series $NH_{4(1-\delta)}D_{4\delta}Br$, regardless of the value of δ, must lie on this dashed Line 1. The error bars are not shown since they are smaller than the data points. The derivation of the experimental errors of $\chi_{Mass}$, $\chi_{Mol}$ and $\delta$ of the parent and mixed samples will be discussed in the last section.

We probed three mixed solids belonging to the $NH_{4(1-\delta)}D_{4\delta}Br$ series each with a different unknown $\delta$. Since the masses of these samples and their magnetizations were measurable, their $\chi_{Mass}$ quantities were obtained. Table S1 lists the molar ratios of the starting solutions and measured $\chi_{Mass}$ quantities. The molar masses of $NH_4Br$ and $ND_4Br$ are 97.942 g/mol and 101.967 g/mol so Eq. 2, for the $NH_{4(1-\delta)}D_{4\delta}Br$ series, has the form $\chi_{Mol} = \chi_{Mass} \cdot [97.942 \text{ g/mol} \cdot (1 - \delta) + 101.967 \text{ g/mol} \cdot \delta]$. The inset of Fig. 4 shows the measured $M_{Mass}$ vs. $H$ curve of one of the samples with unknown $\delta$. The slope of the linear fit yielded $\chi_{Mass} = -5.0495(12)\times10^{-7}$ emu/g-Oe so upon substitution to Eq. 2, Line 2 for this sample was obtained. Figure 4 shows both Line 1 (dashed orange line) and Line 2 (dotted pink line) from which their intersection yielded $\delta = 0.634(25)$ and $\chi_{Mol} = -50.744(61)\times10^{-6}$ cm$^3$/mol. Note that for this mixed sample, the starting solution was a 1:2 ratio of hydrogen to deuterium atoms. The $\chi_{Mass}$ values of the other two samples with starting solutions of different concentrations are listed in Table S1 and their respective $\delta$ were obtained the same way.

To verify the accuracy of the presented approach, $\delta$ of the same set of samples were determined from another method and contrasted. The lattice constants of the $NH_{4(1-\delta)}D_{4\delta}Br$ series are additive[9,27] so the lattice parameters of the pure compounds $NH_4Br$ and $ND_4Br$ can be connected by a straight line (dashed line in Fig. S1). And by measuring the lattice constants of the three mixed samples their $\delta$ values can be derived. This is a common method to determining $\delta$ of mixed inorganic compounds. In the case of the $NH_{4(1-\delta)}D_{4\delta}Br$ series, the samples are cubic at room temperature so there is only one lattice constant $a$.[28] The last column in Table S1 contrasts the obtained $\delta$ from both XRD and magnetic measurements; their differences were all less than 4%.

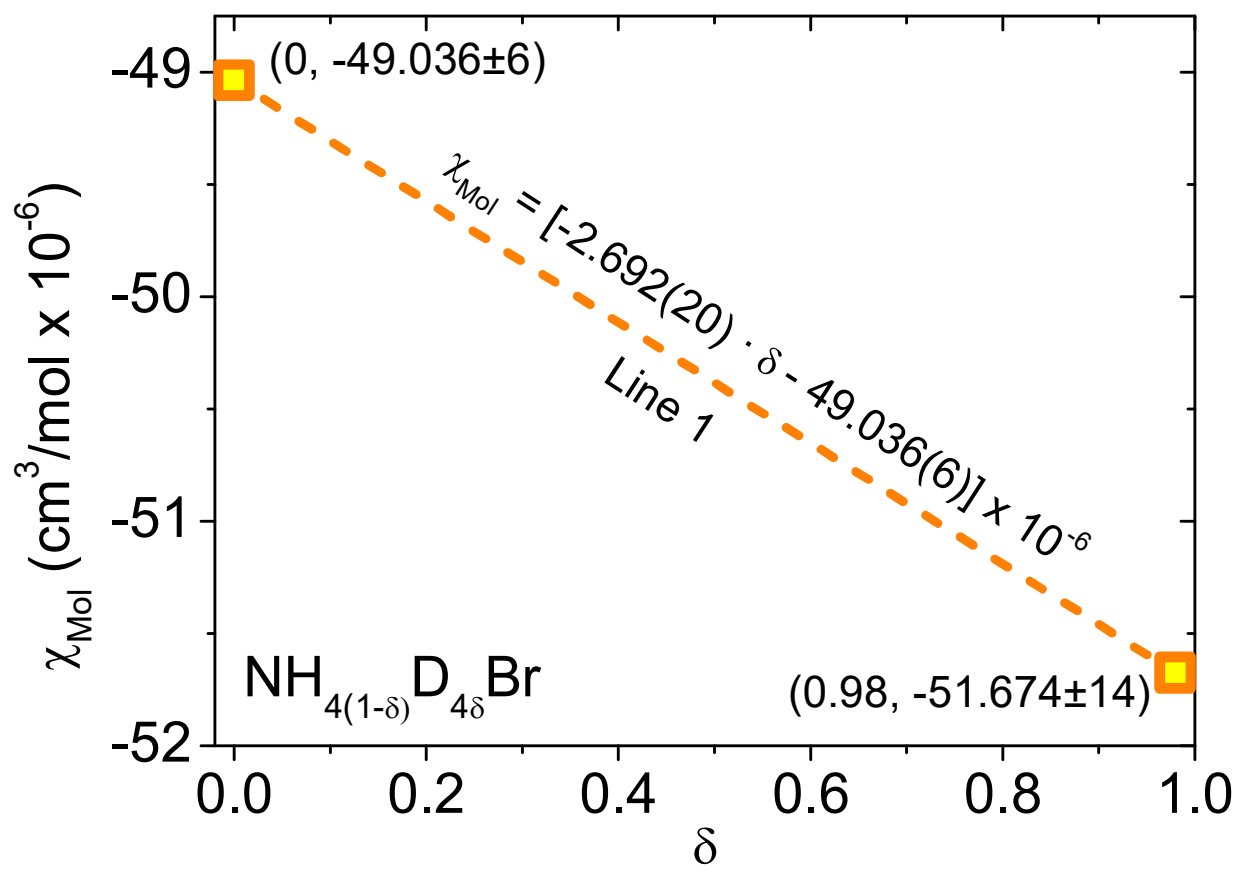

**Fig. 3.** Relationship between sample composition and molar magnetic susceptibility of the $NH_{4(1-\delta)}D_{4\delta}Br$ series, Line 1.

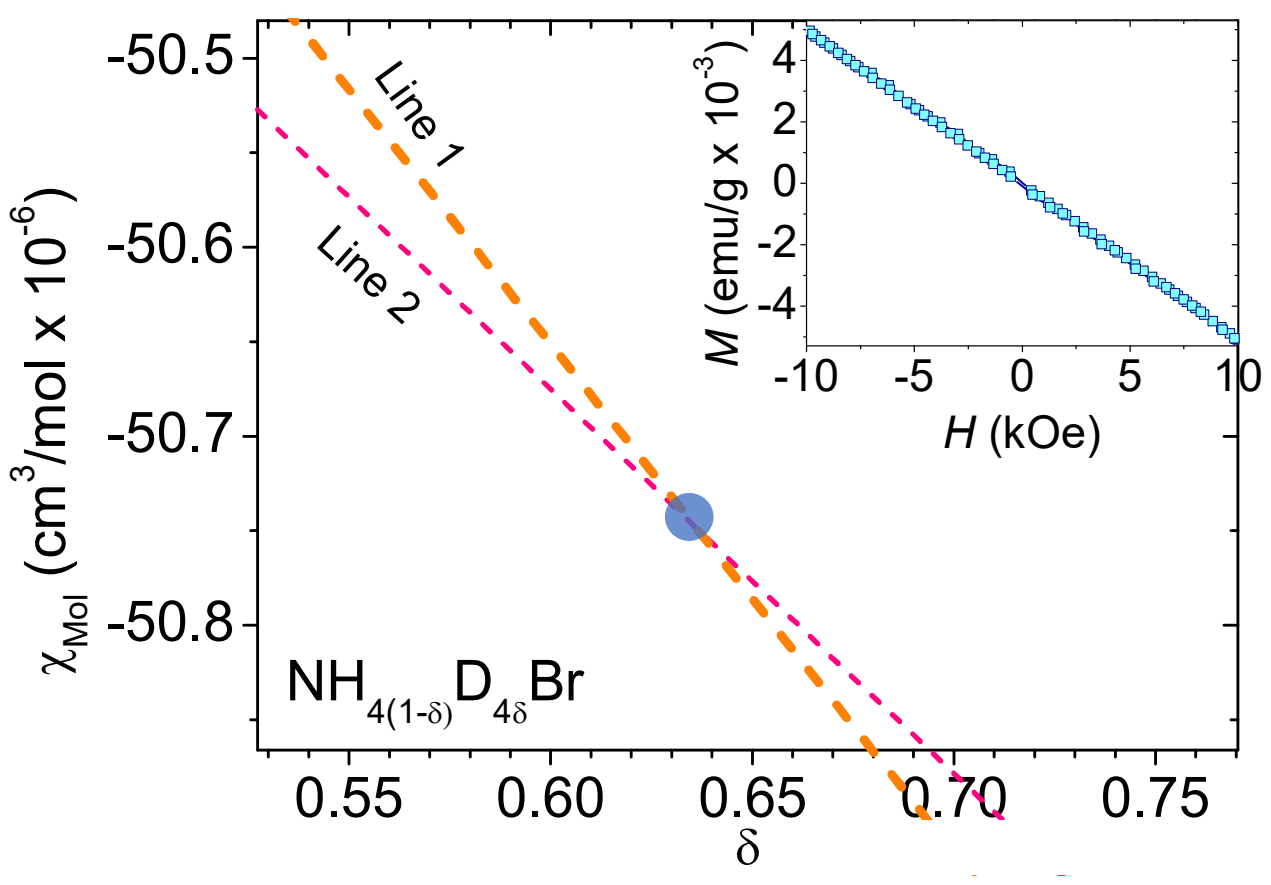

**Fig. 4.** A second equation with the same two unknown variables $\chi_{Mol}$ and δ is available from each mixed sample since their mass susceptibility $\chi_{Mass}$ is different. $\chi_{Mass}$ of each sample is obtained from the slope of the linear fit of their measured *M*(*H*) curves (inset).

The average experimental error of δ obtained for the mixed samples was 0.017, meaning that δ can be resolved to 58 parts, a rather low value since the difference of $\chi_{Mol}$ between $NH_4Br$ and $ND_4Br$ was only $2.692(20)\times10^{-6}$ $cm^3$/mol. However, comparing this to the method of determining δ from lattice constant measurements, in the same range of 0 to 1, δ can only be resolved to 59 parts. This is because the precision of XRD equipment in typical laboratories is 0.0001 Å and the lattice constants of $NH_4Br$ and $ND_4Br$ at room temperature are 4.0659 and 4.0600 (Fig. S1). This example is the extreme case when the difference of $\chi_{Mol}$ between the two parent compounds is small, all of the rest of the examples have $\chi_{Mol}$ differences much larger and most have much better δ resolutions.

The slope of Line 1 is the difference in $\chi_{Mol}$ between $NH_4Br$ and $ND_4Br$, or more specifically, the difference of the $NH_4^+$ and $ND_4^+$ cations in units of $cm^3$/mol. The Pascal's constant of $NH_4^+$ is $-13.3\times10^{-6}$ $cm^3$/mol, so according to our results, that for the $ND_4^+$ cation is $-16.0\times10^{-6}$ $cm^3$/mol; we could not find this value in any existing literature.

### Example 2: $NH_4I_{1-\delta}Br_\delta$

At room temperature, both $NH_4I$ and $NH_4Br$ are cubic, however the former is face-centered of the NaCl type and the latter body-centered of the CsCl type.[28] Despite a large difference in lattice constants, 7.2758 Å for $NH_4I$ and 4.0659 Å for $NH_4Br$,[28] the solutions of these two compounds are completely miscible and its lattice constant is additive up to a Br- concentration of 37%.[9] Above this critical point the lattice constant decreases by over 3 Å and becomes nearly independent of Br- concentration so the $NH_4I_{1-\delta}Br_\delta$ series may be treated to be partially additive. As such, the composition of these mixed solids may only be determined when $\delta < 0.37$ if we were to only work with lattice constants.

Four mixed solids belonging to the $NH_4I_{1-\delta}Br_\delta$ series with starting solutions listed in Table S2 were prepared. The same approach as that in Example 1 was taken. First, Line 1 was constructed from the measured molar susceptibility of the pure compounds $NH_4I$ and $NH_4Br$: $\chi_{Mol} = [14.932(13) \cdot \delta - 63.968(7)]\times10^{-6}$ $cm^3/mol$. Then, Line 2 for each of the mixed solids were obtained from their measured $\chi_{Mass}$ values which had the form of $\chi_{Mol} = \chi_{Mass} \cdot [144.942 \text{ g/mol} \cdot (1 - \delta) + 97.942 \text{ g/mol} \cdot \delta]$. The intersections between Line 1 and Line 2 for each of the four mixed samples are shown in Fig. 5.

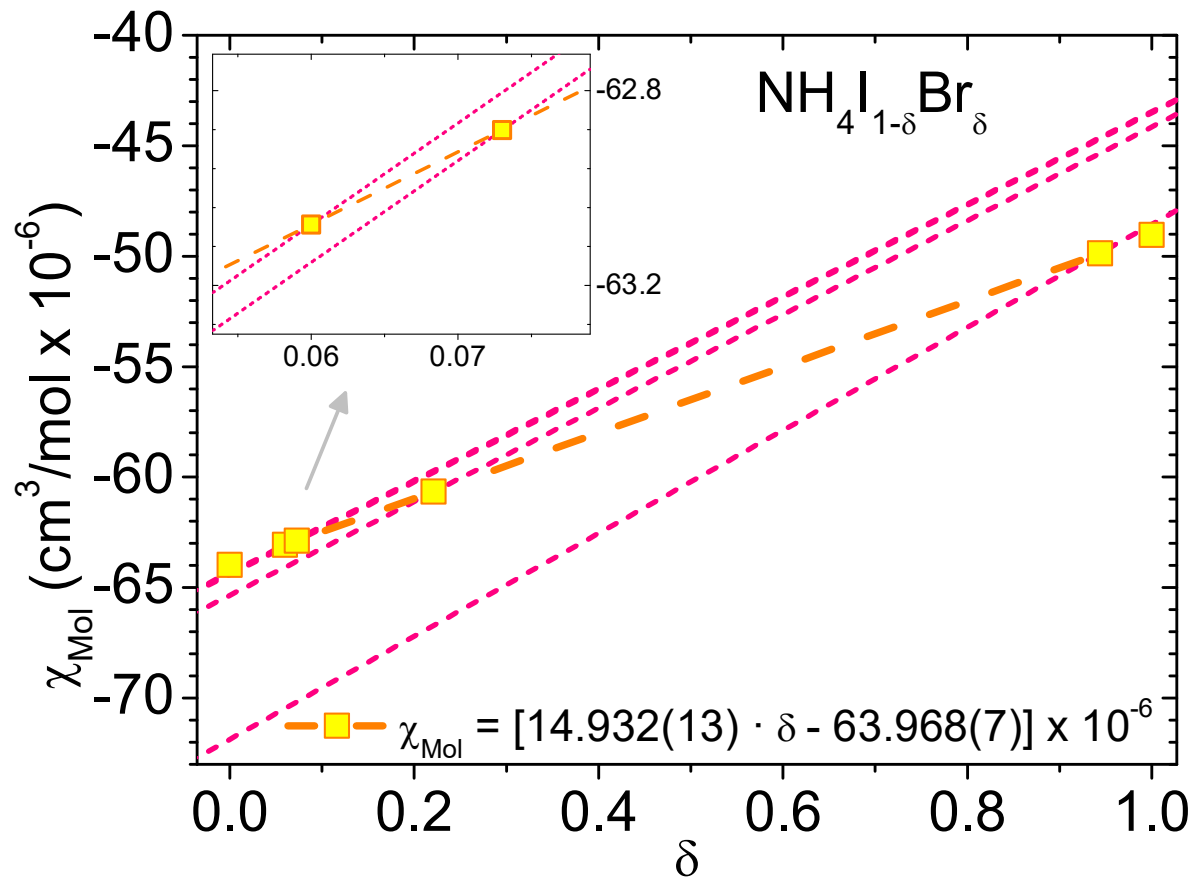

**Fig. 5.** Experimentally obtained values of $\chi_{Mol}$ of $NH_4I$, $NH_4Br$ and those of four mixed solids in the series $NH_4I_{1-\delta}Br_\delta$. The orange dashed line (Line 1) represents the only allowed combinations of $\chi_{Mol}$ and δ for all mixed samples of $NH_4I$ and $NH_4Br$. Each of the four dotted lines (Line 2) represents Eq. 2 of a different mixed solid; their intersections with Line 1 allows for the determination of δ. Inset is an enlarged region showing that the upper dashed line is comprised of two dashed lines.

The lattice constants were also measured for the four mixed solids and plotted as a function of δ in Fig. S2. As a comparison, the data points from Havighurst *et al.*[9] are also plotted in Fig. S2. There is a discrepancy of ~0.03 Å between the two data sets when $\delta < 0.37$ which we attribute to sample purity because the difference was also observed in the undoped sample $NH_4I$. The samples we acquired were 99.999% in purity. If the fitting line used in Ref. 9 is shifted by 0.03 Å to serve as the control measure, then the δ values obtained from Eqs. 1 & 2 only deviate by ~1%.

The inset of Fig. S2 shows the behavior of the lattice constant in the entire $0 < \delta < 1$ range. While the lattice constant of the $NH_4I_{1-\delta}Br_\delta$ series is only partially additive, its molar magnetic susceptibility *is* additive (Fig. 5). This means that a second control method is not needed to check the limits of additivity if the current method were to be employed to determine δ.

The experimental error of δ obtained for the mixed samples in this example were rather low with the 12.8 mg sample having the highest error of 0.006 which translates to δ capable of being resolved to 167 parts.

**Example 3: $(NH_4H_2)_{1-\delta}(ND_4D_2)_\delta PO_4$**

The lattice constant of ammonium dihydrogen phosphate (ADP) $NH_4H_2PO_4$ does not vary linearly with respect to the amount of deuteration[29] because the $D^+$ can replace the $H^+$ of either the ammonium cations or acid protons at different probabilities. As such, lattice constant measurements cannot determine δ of the ADP and deuterated ammonium dihydrogen phosphate (DADP) mixed system $(NH_4H_2)_{1-\delta}(ND_4D_2)_\delta PO_4$ unless another analytical method was used prior to calibrate the lattice constants-to-δ relationship. One way to determine the amount of deuteration of ADP is through the transition temperature of its order-disorder phase transition.[30] The critical temperatures $T_C$ of ADP and DADP are 148 K and 242 K,[31,32] respectively, with δ varying linearly with respect to $T_C$.[33]

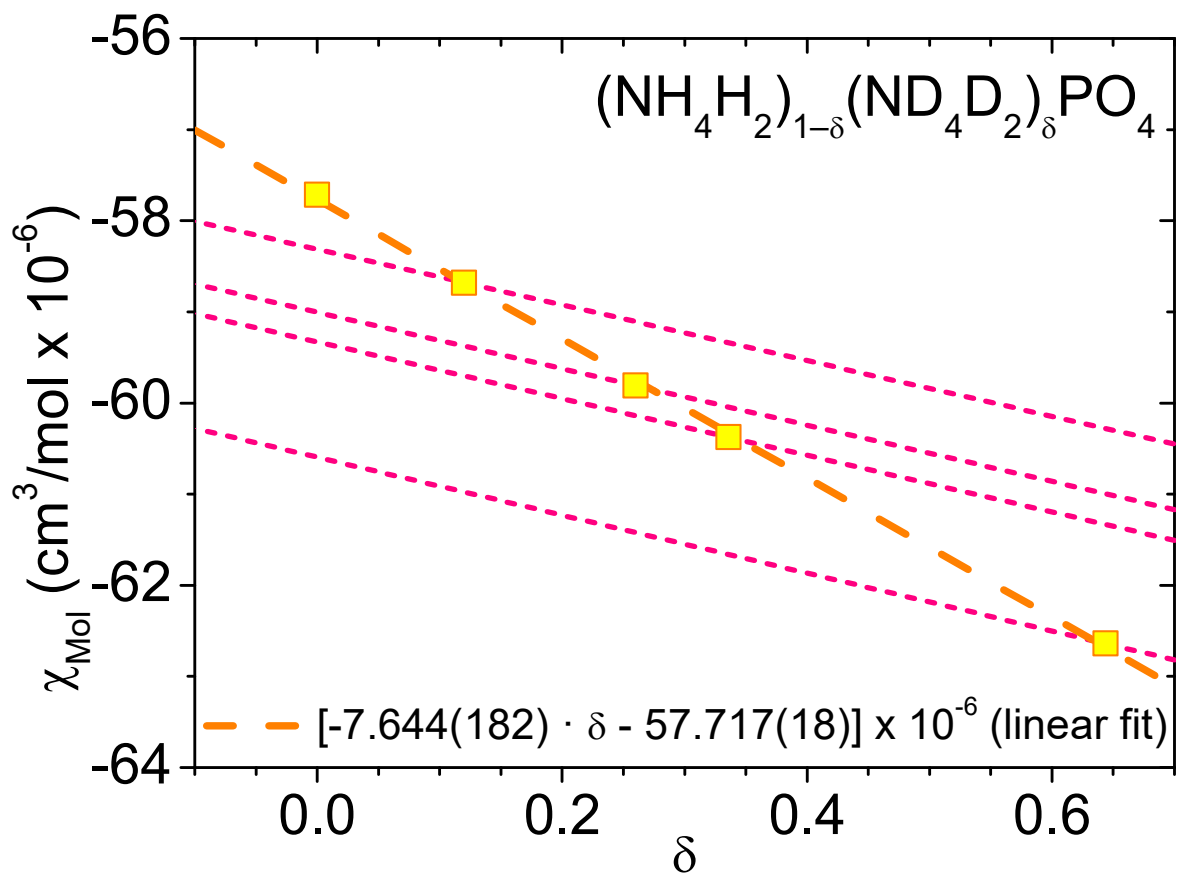

**Fig. 6.** Line 2 (pink dotted) for each of the four mixed solids and their intersections with Line 1 (orange dashed) of the $(NH_4H_2)_{1-\delta}(ND_4D_2)_\delta PO_4$ series. Note that Line 1 was derived from the $T_C$'s of the samples.

The Supplementary Information file describes the process of extracting $T_C$ from the monitoring of the near-static dielectric constant; Table S3 lists the obtained δ for the four mixed solids. For the same four mixed samples as well as pure ADP, their $\chi_{Mass}$ were also measured so their expected $\chi_{Mol}$ values were also obtained according to Eq. 1. The five values of $\chi_{Mol}$ were plotted in Fig. 6 and a linear fit (orange dashed line) was applied to obtain Line 1 for the series $(NH_4H_2)_{1-\delta}(ND_4D_2)_\delta PO_4$. While Line 1 can be established by measuring the $\chi_{Mol}$ of pure ADP and pure DADP which should yield the same Line 1 as shown in Figure 6, we did not take the route of growing a fully deuterated sample for this set of experiments. Instead, δ was first determined through their critical transition temperatures and directly substituted into Eq. 1 to obtain Line 1: $\chi_{Mol} = [-7.644(182) \cdot \delta - 57.717(18)]\times10^{-6}$ cm$^3$/mol. Suppose now that δ was unknown for the four mixed solids, hence, Eq. 2 must be employed and each mixed sample will have a Line 2. Figure 6 also shows the four Line 2's of the mixed samples; their parameters $\chi_{Mass}$ and intersections with Line 1 are listed in Table S3. The close values of δ obtained from the dielectric and magnetic methods were rather gratifying. Furthermore, Figure 6 shows that even in the case of more complex mixed diamagnetic systems that are not binary compounds, their magnetic susceptibility is also additive.

**Example 4: $C_{48}H_{22+6\delta}Br_{6(1-\delta)}O_{32}Zr_6$**

MOFs that are not diamagnetic usually behave paramagnetically because the magnetic ions within dimers or trimers are far apart from other dimers or trimers of the same species thereby limiting intermolecular antiferro- or ferro-magnetic interactions. Since ***H*** and ***m*** are also linear in paramagnetic systems, our presented method should also be applicable. The main difference will be that $\chi_{Mol}$ is positive rather than negative. The magnitude of $\chi_{Mol}$ of paramagnetic systems with only one unpaired electron is already at least 1 to 2 orders of magnitude larger than their diamagnetic contributions so their quantitative analyses using Eqs. 1 & 2 should yield even more accurate results. In the cases when the electron spin configurations of the parent MOFs are identical, such as when the only difference is the replacement of a ligand,[34-36] then the problem of determining the sample composition of its mixed compounds using our measure boils down to working with their diamagnetic corrections. To lay down the foundation of testing MOFs, we selected $Zr^{4+}$-based MOFs UiO-66 and Uio-66-Br.

Figure 7 shows $\chi_{Mol}$ vs. δ of the series $C_{48}H_{22+6\delta}Br_{6(1-\delta)}O_{32}Zr_6$. The mixed samples were directly taken from a dry mixture of $C_{48}H_{28}O_{32}Zr_6$ (UiO-66) and $C_{48}H_{22}Br_6O_{32}Zr_6$ (UiO-66-Br) and the control values of δ were obtained from the mass ratio of the two pure compounds. UiO-66 and UiO-66-Br were assigned as A' and A so Line 1 for this series was $\chi_{Mol} = [-23.41(99) \cdot \delta - 765.11(66)]\times10^{-6}$ cm$^3$/mol and Lines 2 of the mixed samples had the form $\chi_{Mol} = \chi_{Mass} \cdot [2137.44 \text{ g/mol} \cdot (1 - \delta) + 1664.06 \text{ g/mol} \cdot \delta]$. The slope of Line 1 is in good agreement with $24.6\times10^{-6}$ cm$^3$/mol, the expected quantity of replacing six C–H bonds with six C–Br bonds according to their Pascal's constants of 0 and $+4.1\times10^{-6}$ cm$^3$/mol,[26] respectively. The values of $\chi_{Mass}$ and obtained data points of $\chi_{Mol}$ and δ from the intersections of Line 1 and Lines 2 are listed in Table S4. The difference between δ obtained from the magnetic method and from that of the control method were on average 0.95%. The largest experimental error obtained for δ was 0.008 for the 10.4 mg mixed sample which resolved δ to 125 parts. From these two statistical values, we expect the presented method to be highly useful in determining ligand ratios of mixed MOFs in powdered form because employing inductively coupled plasma mass spectroscopy (ICP-MS) to obtain δ is challenging when the two parent compounds do not have an element that the other does not have.

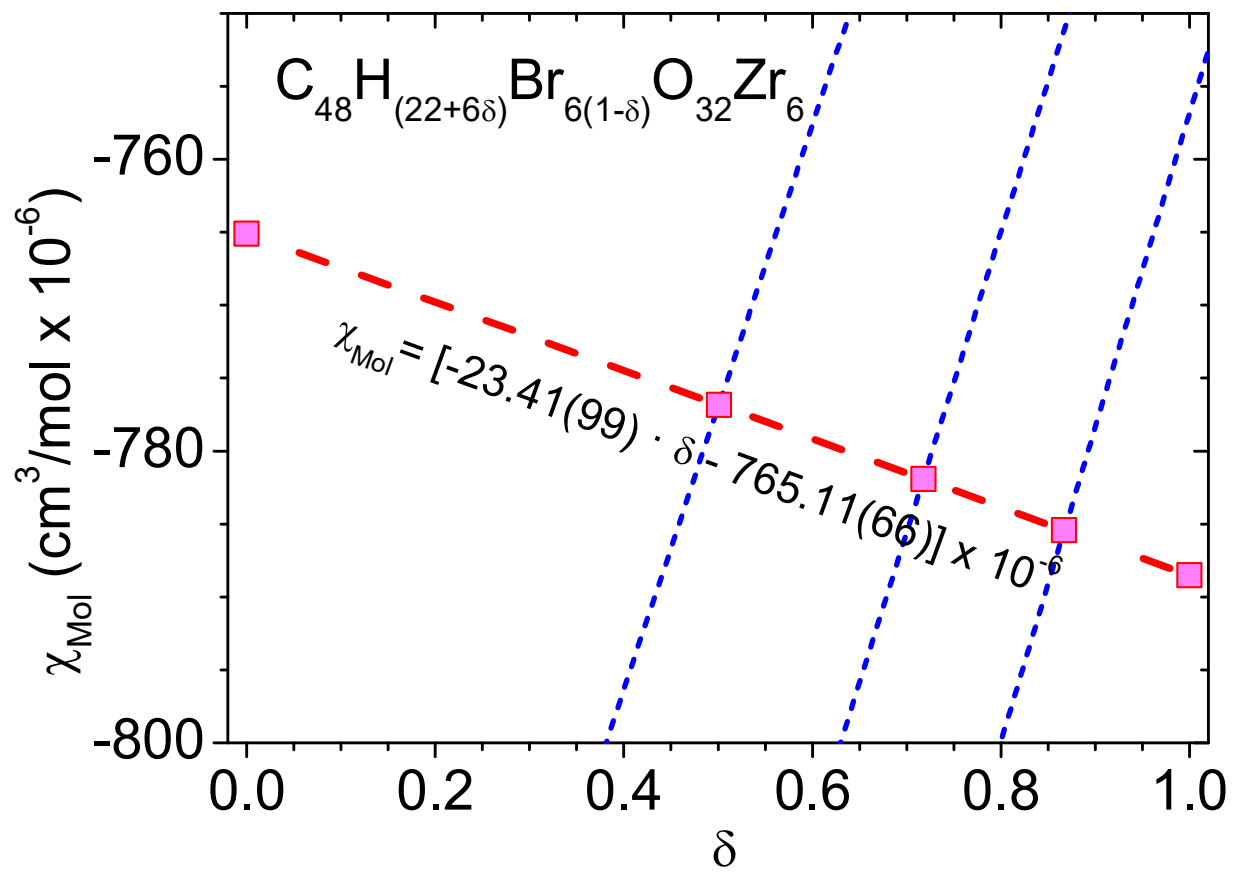

**Fig. 7.** Experimentally obtained values of $\chi_{Mol}$ and δ of the $C_{48}H_{22+6\delta}Br_{6(1-\delta)}O_{32}Zr_6$ metal organic framework system. The five data points and slopes ($\chi_{Mass}$) of Lines 2 are listed in Table S4.

**Example 5: [creatine]$_{1-\delta}$[D-glucose]$_{\delta}$**

D-Glucose can crystallize into two polymorphs, α and β, in ambient conditions to give rise to two crystal structures.[37,38] To complicate matters, the two phases in glucose interconvert back and forth with time. The main difference between the α and β phases is the direction of one of its –OH bonds so their magnetic susceptibility should also be the same. Figure 8 shows the obtained $\chi_{Mol}$ of creatine and D-glucose (δ = 0 and 1, respectively) as well as the extracted δ of three of their mixed compounds. Line 1 for this series was $\chi_{Mol} = [-32.03(21) \cdot \delta - 74.950(18)]\times10^{-6}$ cm$^3$/mol and Lines 2 had the form $\chi_{Mol} = \chi_{Mass} \cdot [131.13 \text{ g/mol} \cdot (1 - \delta) + 180.16 \text{ g/mol} \cdot \delta]$. The solved values of $\chi_{Mol}$, δ and their respective $\chi_{Mass}$ values are listed in Table S5. With the use of Eqs. 1 and 2, the ratio between the α and β phases of glucose cannot be discerned, however, the ratio between glucose (regardless of its α and β concentrations), when mixed with another compound, is discernable. This is complimentary to working with lattice parameters to determining chemical composition because XRD measurements have the opposite problem.

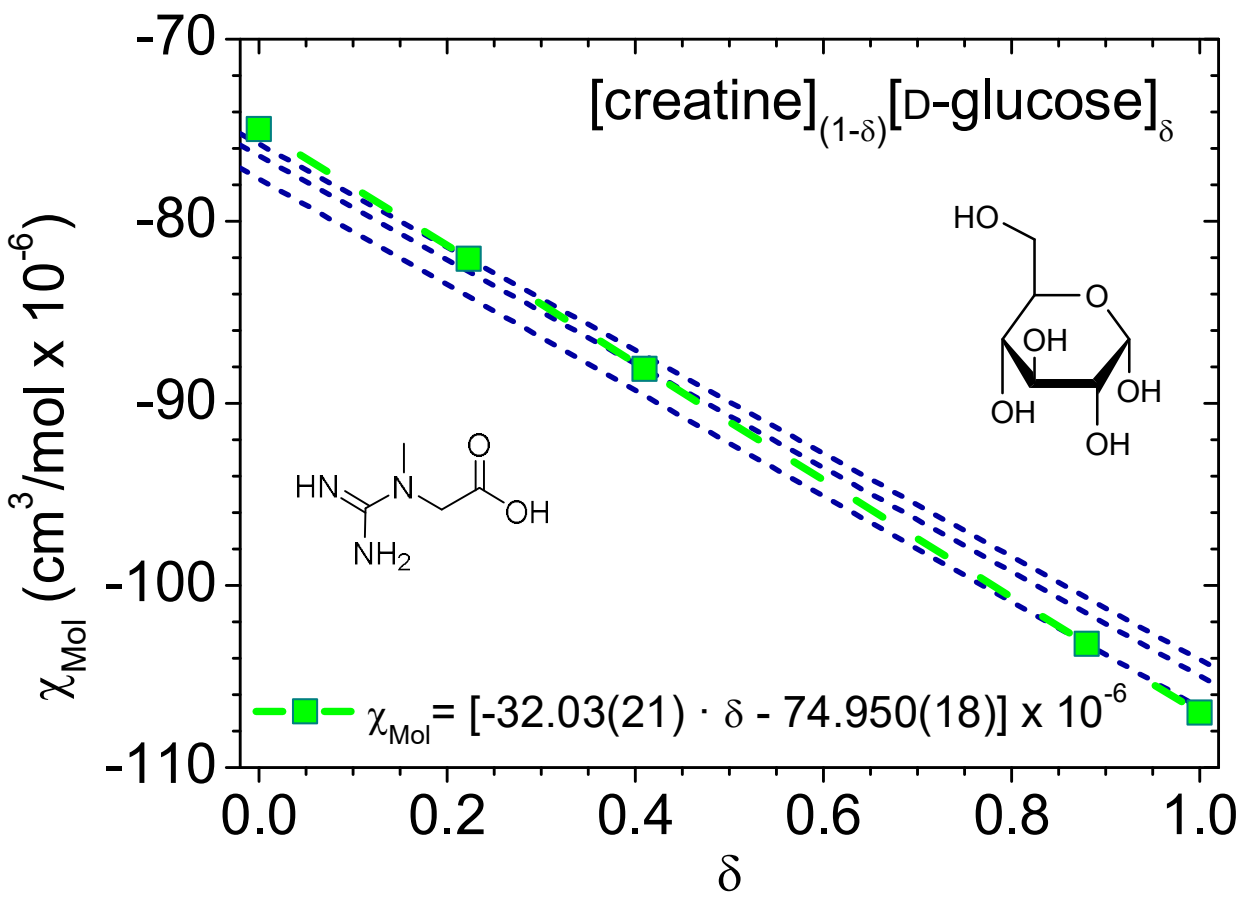

**Fig. 8.** $\chi_{Mol}$ and δ of five samples of the [creatine]$_{(1-\delta)}$[D-glucose]$_\delta$ series.

### Example 6: [L-glutamic acid]$_{1-\delta}$[L-leucine]$_\delta$

L-Glutamic acid is similar to D-glucose in the sense that two polymorphs of the same compound are stable at ambient conditions.[39,40] L-Leucine as well as many other amino acids are increasingly being employed as dopants to enhance the non-linear optical properties of ferroelectric crystals[41,42] so an accurate determination of chemical composition is of high importance. Figure 9 shows the experimentally obtained $\chi_{Mol}$ and δ for the series [L-glutamic acid]$_{1-\delta}$[L-leucine]$_\delta$ for when δ was 0, 1 and three other arbitrary values. The obtained Lines 1 & 2 for this series were, respectively, $\chi_{Mol} = [-12.957(41) \cdot \delta - 81.529(21)]\times10^{-6}$ $cm^3$/mol and $\chi_{Mol} = \chi_{Mass} \cdot [147.13 \text{ g/mol} \cdot (1 - \delta) + 131.17 \text{ g/mol} \cdot \delta]$. The experimental error obtained for δ for the mixed samples of this example was the smallest of all; δ was resolved to 333 parts with an average sample mass of only 12.76 mg (Table S6).

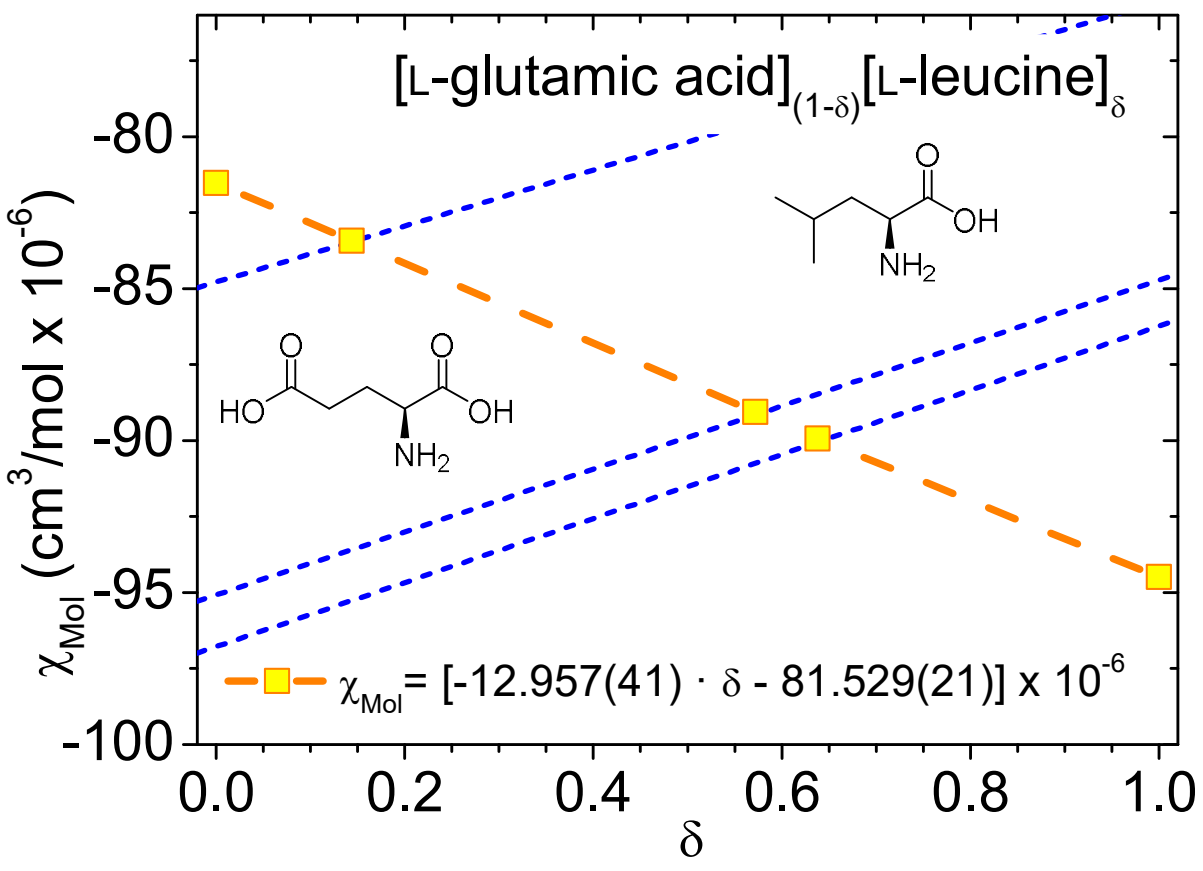

**Fig. 9.** $\chi_{Mol}$ and δ of five samples of the [L-glutamic]$_{(1-\delta)}$[L-leucine]$_\delta$ series.

**Example 7: [terephthalic acid]$_{1-\delta}$[trimesic acid]$_{\delta}$**

Terephthalic acid and trimesic acid are commonly used reagents in most laboratories; the former often serve as precursors to plasticizers and linkers to MOFs. The main difference between these two compounds is the replacement of a C–H bond of the benzene ring by a C–COOH bond. Figure 10 shows Eq. 1 possessing the form $\chi_{Mol} = [-10.734(47) \cdot \delta - 86.434(14)] \times 10^{-6}$ cm$^3$/mol and Eq. 2 of the mixed compounds with the form $\chi_{Mol} = \chi_{Mass} \cdot [166.13\ \text{g/mol} \cdot (1 - \delta) + 210.14\ \text{g/mol} \cdot \delta]$. In this example, deviation of δ from its control values were the smallest (Table S7) when compared to all other examples and the experimental errors of δ were all less than 0.007 allowing it to be resolved to 143 parts.

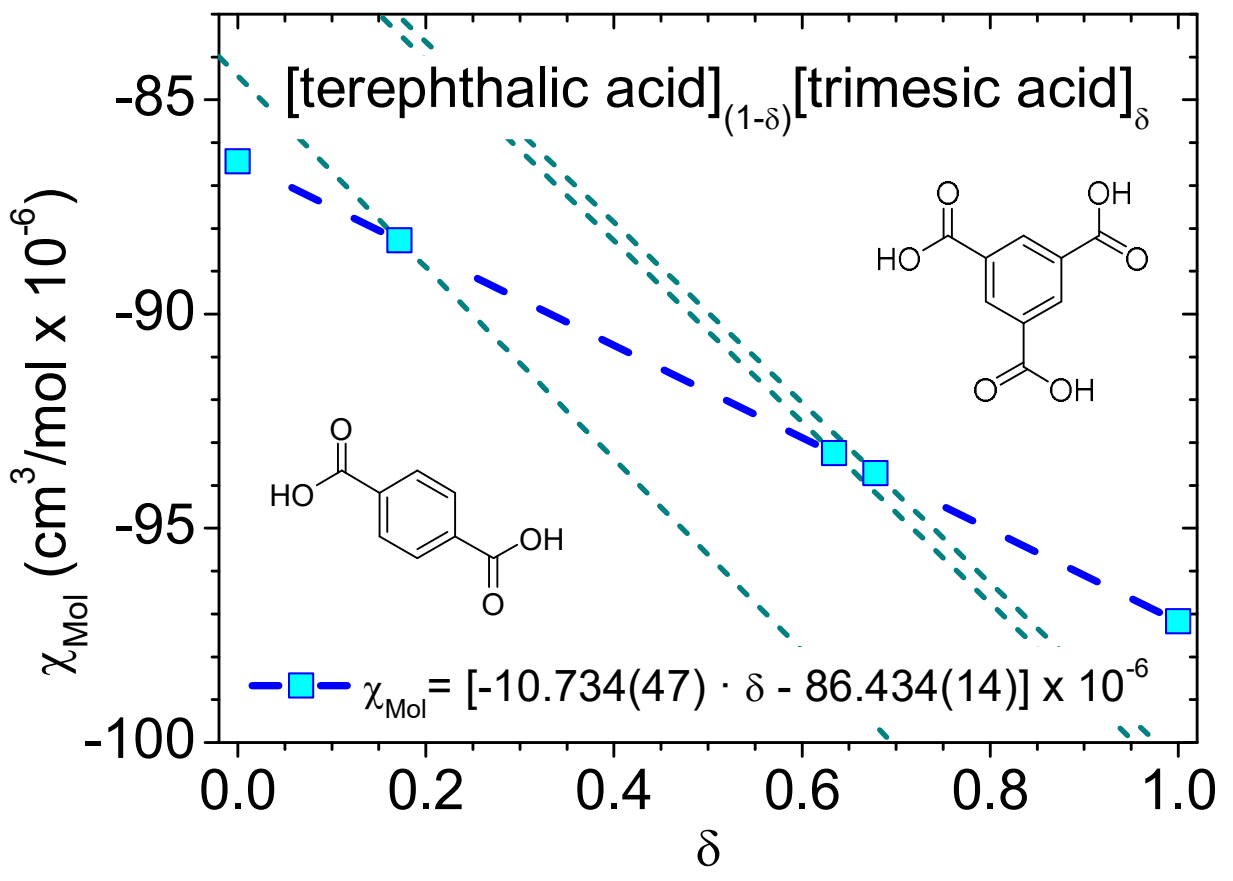

**Fig. 10.** $\chi_{Mol}$ and δ of five samples of the [terephthalic acid]$_{(1-\delta)}$[trimesic acid]$_{\delta}$ series.

**Example 8: [triphenylphosphine]$_{1-\delta}$[*p*-terphenyl]$_{\delta}$**

Triphenylphosphine ($PPh_3$) is a common ligand used in the synthesis of many organometallic compounds and *p*-terphenyl differs mainly by a P atom so these two compounds were selected to further show the applicability of the presented method. Figure 11 shows $\chi_{Mol}$ and δ of $PPh_3$, *p*-terphenyl and three of their mixed compounds. The equations obtained were $\chi_{Mol} = [19.15(8) \cdot \delta - 178.54(4)] \times 10^{-6}$ cm$^3$/mol serving as Line 1 and $\chi_{Mol} = \chi_{Mass} \cdot [262.3\ \text{g/mol} \cdot (1 - \delta) + 230.3\ \text{g/mol} \cdot \delta]$ as Lines 2 for each of the mixed samples. The set of mixed samples of this series exhibited the largest experimental errors in δ because the slopes of Line 1 and Lines 2 were very similar (Table S8). Omission of this example would have allowed us to reduce the average experimental error of all mixed samples from 1.2% down to 0.86%, *i.e.* obtain an average resolution of 117 parts instead of 83, however, there will be extreme cases such as these so we want to provide a full picture of the presented method.

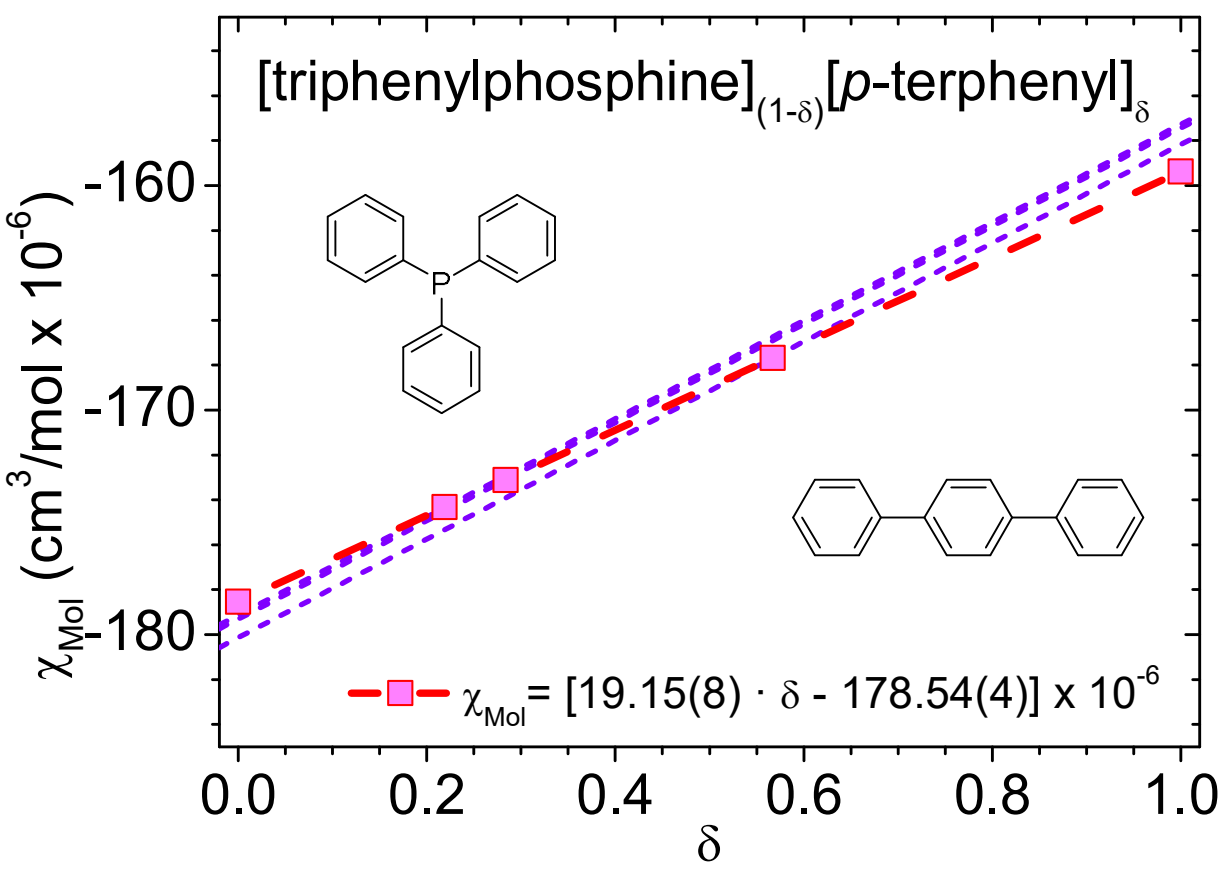

**Fig. 11.** $\chi_{Mol}$ and δ of five samples of the [triphenylphosphine]$_{(1-\delta)}$[$p$-terphenyl]$_\delta$ series.

**Error analysis:**

The noise level of the measured magnetization ***m*** of the magnetometer (PPMS) is rated as $2\times10^{-7}$ emu at ±10 kOe. This is the case if only one point was measured, however, in each of our experiments we performed a 10 kOe → –10 kOe → 10 kOe scan averaging at least 100 points (inset of Fig. 4) so the noise level was reduced by 1 order of magnitude. The typical signal measured at 10 kOe in the current experiments ranges from $-2\times10^{-4}$ to $-2\times10^{-5}$ emu (average sample mass of 16.6 mg) so the measurements have signal-to-noise ratios of 1,000–10,000. As an example, the averaged ***m*** at 10 kOe of the parent samples UiO-66-Br (δ = 0) and UiO-66 (δ = 1) were $-2.32671\times10^{-5}$ emu and $-4.73856\times10^{-5}$ emu, respectively, so they had signal-to-noise ratios of 1,163 and 2,369. From such, their $\chi_{Mass}$ along with their experimental error limits were $-0.35796(31)\times10^{-6}$ emu/g and $-0.47386(20)\times10^{-6}$ emu/g; their $\chi_{Mol}$ were, respectively, $-765.11(66)\times10^{-6}$ cm$^3$/mol and $-788.52(33)\times10^{-6}$ cm$^3$/mol. To obtain the experimental error limits of δ and $\chi_{Mol}$ of the mixed samples, Lines 1 and 2, along with their maximum possible errors, need to be plotted and analyzed. Line 1 of Example 4 along with its margin of error are shown in Fig. 12 in red. Line 1 has the equation $\chi_{Mol} = [-23.41 \cdot \delta - 765.11]\times10^{-6}$ cm$^3$/mol (red dashed line) and the limit of its error margins have the form $\chi_{Mol} = [-22.42 \cdot \delta - 764.45]\times10^{-6}$ cm$^3$/mol and $\chi_{Mol} = [-24.40 \cdot \delta - 765.77]\times10^{-6}$ cm$^3$/mol (red solid lines). The δ = 0.718 sample was selected for this example, its $\chi_{Mass}$ was $-0.43495(18)\times10^{-6}$ emu/g so Line 2 has the form of $\chi_{Mol} = -0.43495\times10^{-6}$ emu/g · [2137.44 g/mol · (1 – δ) + 1664.06 g/mol · δ] (blue dashed line) and its error margins have the form $\chi_{Mol} = -0.43477\times10^{-6}$ emu/g · [2137.44 g/mol · (1 – δ) + 1664.06 g/mol · δ] and $\chi_{Mol} = -0.43513\times10^{-6}$ emu/g · [2137.44 g/mol · (1 – δ) + 1664.06 g/mol · δ] (blue solid lines). The four orange dots represent the largest possible errors when working with both Eqs. 1 and 2. The orange dots having the largest *x*- or *y*-components were taken as the error margins of δ and $\chi_{Mol}$. Hence, for this particular mixed sample, $\delta = 0.718(7)$ and $\chi_{Mol} = 781.90(125)$ so in the range of $0 < \delta < 1$, $\delta$ was resolved to 143 parts. The average mass and resolutions of all of the 26 mixed samples in the studied eight examples were 16.6 mg and 83 parts, respectively. The resolution of $\delta$ scales with the amount of sample so for resolutions of 100 parts, only ~20 mg of sample in powdered form is needed. As a comparison, NMR measurements employed for the same characterization method can typically determine $\delta$ to resolutions of around 100 parts; however, a significant amount of sample size is needed. For instance, at least 1 $cm^3$ of sample is required for solid-state NMR measurements.[43] In the current experiments, a 12-fold increase of the sample masses can allow $\delta$ to be potentially resolved to $>10^3$ parts. Furthermore, employing a SQuID (Superconducting Quantum Interference Device) magnetometer, which has noise levels one order of magnitude lower than the currently employed vibrating sample magnetometer, can render $\delta$ to be resolved to higher accuracy.

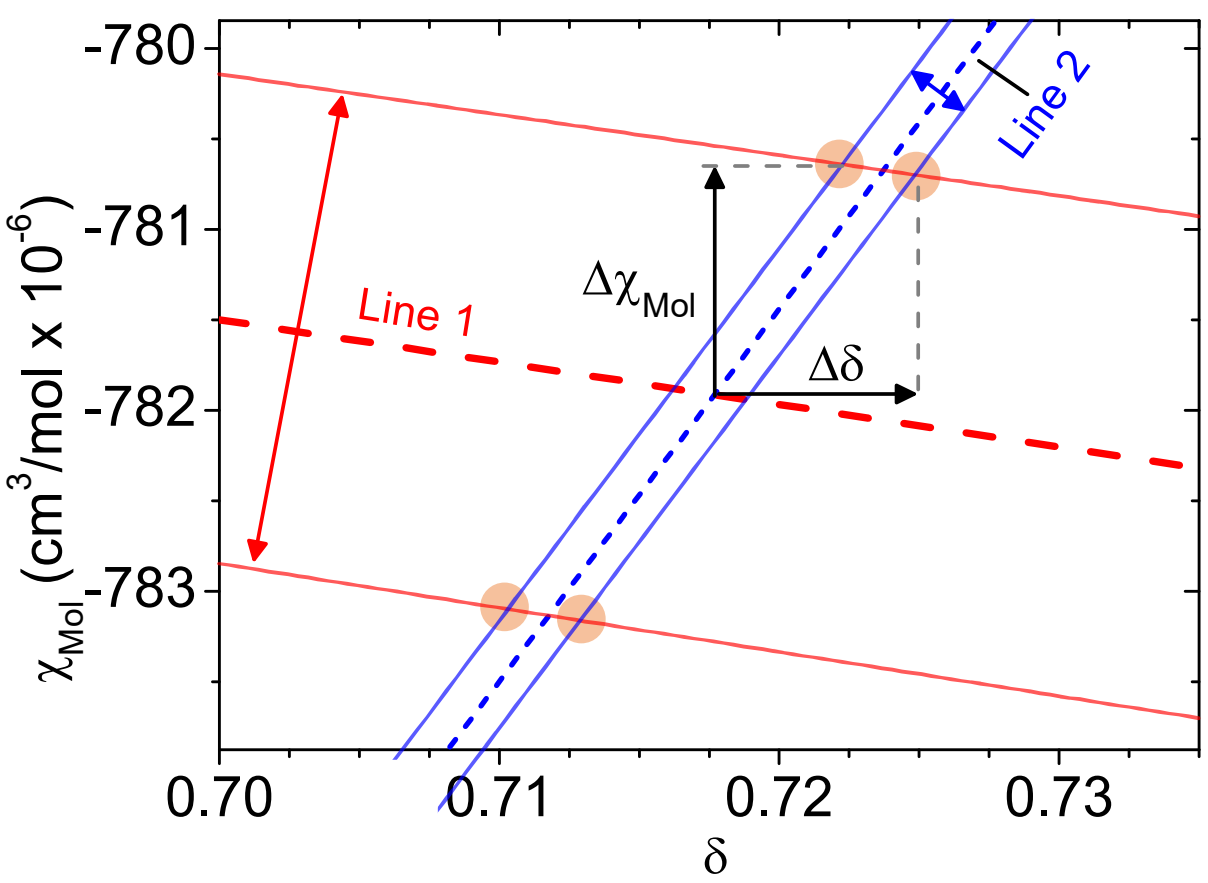

**Fig. 12.** Line 1 (dashed red line) along with its error margin (solid red lines) of the $C_{48}H_{22+6\delta}Br_{6(1-\delta)}O_{32}Zr_6$ system discussed in Example 4. Line 2 (dashed blue line) of the $\delta = 0.718$ mixed sample in the same system along with its error margin (solid blue lines). Orange dots represent the largest possible error when employing both Eqs. 1 and 2. The error margins of $\delta$ and $\chi_{Mol}$ were taken as the largest *x*- and *y*-components bounded by the orange dots.

### Precautions and limitations

During the measurement of $\boldsymbol{m}$, the samples need to be attached onto a sample rod. Usually, a form of adhesive is used to accomplish this. In our case, we used GE varnish; the mass needed to firmly secure the sample ranged from 12-16% of that of the sample. These adhesives also have a diamagnetic response so their $\chi_{Mass}$ need to be measured first to then serve as a background to be subtracted during the sample measurements. Hence, during each measurement of the sample, the mass of the adhesive also needs to be weighed.

It is highly recommended to experimentally measure $\chi_{Mol}$ of the parent compounds instead of just obtaining a theoretical approximation of $\chi_{Mol}$ through addition of Pascal's constants because the experimental values are almost always different due to sample purity. Table S9 lists the experimentally obtained $\chi_{Mol}$ of all of the parent compounds along with their expected values from the addition of Pascal's constants.

Since covalent bonds are directional, the diamagnetic response of a molecule is different depending on the direction of $\boldsymbol{H}$. Hence, $\chi_{mol}$ of a compound that is in single crystalline form is anisotropic. As such, bulk crystalline samples should be grinded into powder and compressed into disc pellets prior to the measurements.

The current approach is not suitable for measuring diamagnetic liquids as most do not obey the simple mixture law. Molten mixtures often undergo chemical and/or physical changes[44,45] so the molecular composition and magnetic susceptibility may not exhibit a linear relationship. The magnetic susceptibility is also very sensitive to liquids dissolving gasses such as oxygen[46] so the current approach may not be suitable to determining the doping or isotope concentrations of porous nanosized materials.

## Conclusion

A method of determining the chemical composition of diamagnetic mixed solids was presented. The trick was to employ two linear equations with the same two unknown variables to simultaneously obtain the molar magnetic susceptibility and doping concentration of a mixed solid. The first (and second) equations use the additive nature of the molar magnetic susceptibility (and molar masses) of the pure compounds. Eight examples were provided to show the feasibility of the presented concept. With experimental errors limited to $\leq$1.2% and only small quantities of powdered samples needed, the presented method should be a highly attractive alternative for determining the chemical composition of mixed solids encompassing a wide range of diamagnetic compounds such as pharmaceutical drugs, composites, polymers, *etc*.

## Author Contributions

M. M. Z. grew the inorganic crystals, performed the magnetic measurements, analyzed data and plotted the figures. Y. Y. was responsible for the organic compounds, XRD characterization and helped with the graphics. N. D. helped with the dielectric constant measurements and sample preparation. Y. Y. Z. assisted with the magnetic measurements. P. R. helped supervise the project and suggested the use of the metal organic framework series and organic compounds. F. Y. designed the project, supervised the experiments and drafted the manuscript.

## References

1. P. Jia, Y. Li, Z. Zheng, Z. Li and L. Cao, *J. Ally. Comp*., 2022, **890**, 161799.
2. S. Du, L. Chen, C. Men, H. Ji, T. Su and Z. Qin, *J. Ally. Comp*., **2023**, 955, 170265.
3. R. J. Spranger, H. Kirchhain, T. M. F. Restle, J. V. Dums, A. J. Karttunen, L. van Wüllen and T. F. Fässler, *J. Phys. Chem. C*, **2023**, 127, 1622.
4. J. Park, K. M. Griffith and Y. Duan, *J. Phys. Chem. C*, **2022**, 126, 8832.
5. X. Li, Y. Xu, C. Zhao, D. Wu, L. Wang, M. Zheng, X. Han, S. Zhang, J. Yue, B. Xiao, W. Xiao, L. Wang, T. Mei, M. Gu, J. Liang and X. Sun, *Angew. Chem. Int. Ed*., **2023**, 62, e202306433.
6. M. Pokora, A. Paneth and P. Paneth, *J. Phys. Chem. Lett*., **2023**, 14, 3735.
7. Y. Y. Xu, L. Meng, M. M. Zhao, C. X. Peng and F. Yen, *J. Ally. Comp*., **2023**, 960, 170685.
8. L. Vegard, *Zeitschrift für Physik*, **1921**, 5, 17.
9. R. J. Havighurst, E. Mack Jr. and F. C. Blake, *J. Am. Chem, Soc*., **1925**, 47, 29.
10. A. Labiche, A. Malandain, M. Molins, F. Taran and D. Audisio, *Angew. Chem. Int. Ed*., **2023**, 62, e202303535.
11. M. H. Thiemens and M. Lin, *Angew. Chem. Int. Ed*., **2019**, 58, 6826.
12. H. Hu, Q. Li, W. Zhang, S. Wu, J. Song, K. Liu, X. Zou, H. Pang, M. Yan and S. Hou, *Geophys. Res. Lett*., **2022**, 49, e2022GL098368.
13. D. Schwarz, G. Henriques, E. Rojo-Wiechel, S. Klare, R. Mika, S. Höthker, J. H. Schacht, N. Schmickler and A. Gansäuer, *Angew. Chem. Int. Ed*., **2021**, 61, e202114198.
14. M. Shao, J. Keum, J. Chen, Y. He, W. Chen, J. F. Browning, J. Jakowski, B. G. Sumpter, I. N. Ivanov, Y. -Z. Ma, C. M. Rouleau, S. C. Smith, D. B. Geohegan, K. Hong and K. Xiao, *Nature Comm*., **2014**, 5, 3180.
15. L. Li, J. Jakowski, C. Do and K. Hong, *Macromolecules*, **2021**, 54, 3555.
16. M. M. Zhao, L. Meng, Y. Y. Xu, N. Du and F. Yen, *J. Phys. Chem. C*, **2023**, 127, 20951.
17. S. S. Zhao, H. Zhang, L. Wang, L. Chen and Z. Xie, *J. Mater. Chem. C*, **2018**, 6, 11701.
18. S. Dou, J. Song, S. Xi, Y. Du, J. Wang, Z. F. Huang, Z. J. Xu and X. Wang, *Angew. Chem. Int. Ed*., **2019**, 58,4041.
19. M. Lightowler, S. Li, X. Ou, X. Zou, M. Lu and H. Xu, *Angew. Chem. Int. Ed*., **2021**, 61, e202114985.
20. S. V. Shishkina, A. M. Shaposhnik, V. V. Dyakonenko, V. M. Baumer, V. V. Rudiuk, I. B. Yanchuk and I. A. Levandovskiy, *ACS Omega*, **2023**, 8, 20661.
21. P. Pascal, *Ann. Chim. Phys*., **1910**, 19, 5.
22. W.R. Myers, *Rev. Mod. Phys*., **1952**, 24, 15.
23. J. Banhart, H. Ebert, J. Voitlander and H. Winter, *J. Mag. Mag. Mat*., **1986**, 61, 221.
24. L. Pauling, *J. Chem. Phys*., **1936**, 4, 673.

25. H. J. Dauben Jr., J. D. Wilson and J. L. Laity, *J. Am. Chem. Soc.*, **1968**, 90, 811.
26. G. A. Bain and J. F. Berry, *J. Chem. Edu.*, **2008**, 85, 532.
27. N. P. Funnel, C. L. Bull, S. Hull and C. Ridley, *J. Phys.: Condens. Matter*, **2022**, 34, 325401.
28. H. A. Levy and S. W. Peterson, *J. Am. Chem. Soc.*, **1953**, 75, 1536.
29. B. Liu, L. Zhu, F. Liu, X. Sun, X. Chen, L. Xie, Y. Xia, G. Sun and X. Ju, *Chin. Phys. B*, **2019**, 28, 016103.
30. L. Meng, C. He, W. Ji and F. Yen, *J. Phys. Chem. Lett.*, **2020**, 11, 8297.
31. B. T. Matthias, *Phys. Rev.*, **1952**, 85, 141.
32. W. P. Mason and B. T. Matthias, *Phys. Rev.*, **1952**, 88, 477.
33. S. R. Gough, J. A. Ripmeester, N. S. Dalal and A. H. Reddoch, *J. Phys. Chem.*, **1979**, 83, 664.
34. S. Kim, K. W. Dawson, B. S. Gelfand, J. M. Taylor and G. K. H. Shumizu, *J. Am. Chem. Soc.*, **2013**, 135, 963.
35. G. C. Shearer, J. G. Vitillo, S. Bordiga, S. Svelle, U. Olsbye and K. P. Lillerud, *Chem. Mat.*, **2016**, 28, 7190.
36. J. Edzards, H.D. Saßnick, A.G. Buzanich, A. M. Valencia, F. Emmerling, S. Beyer and C. Cocchi, *J. Phys. Chem. C*, **2023**, 127, 21456.
37. T. R. R. McDonald and C. A. Beevers, *Acta Cryst.*, **1952**, 5, 654.
38. W. G. Ferrier, *Acta Cryst.*, **1960**, 13, 678.
39. Y. Mo, L. Dang and H. Wei, *Ind. Eng. Chem. Res.*, **2011**, 50, 10385.
40. T. C. Lai, S. Ferguson, L. Palmer, B. L. Trout and A. S. Myerson, *Org. Process Res. Dev.*, **2014**, 18, 1382.
41. C. M. Raghavan, R. Sankar, R. Mohan Kumar and R. Jayavel, *Mat. Res. Bull.*, **2008**, 43, 305.
42. P. Kumaresan, S. Moorthy Babu and P. M. Anbarasan, *Opt. Mat.*, **2008**, 30, 1361.
43. A. J. Howarth, A. W. Peters, N. A. Vermeulen, T. C. Wang, J. T. Hupp and O. K. Farha, *Chem. Mat.*, **2017**, 29, 26.
44. V. C. G. Trew and J. F. Spencer, *Proc. Roy. Soc. London A*, **1931**, 131, 209.
45. Y. Shirakawa, S. Tamaki, H. Okazaki and M. Azuma, *J. Phys. Soc. Jpn.*, **1993**, 63, 544.
46. K. Sueoka, *Bull. Chem. Soc. Jpn.*, **1994**, 67, 978.

# Determining the chemical composition of diamagnetic mixed solids via measurements of the magnetic susceptibility

Miao Miao Zhao,† Yang Yang,† Na Du, Yu Ying Zhu, Peng Ren,* Fei Yen*

*School of Science, Harbin Institute of Technology,*
*Shenzhen, University Town, Shenzhen, Guangdong 518055 P. R. China.*

†These authors contributed equally;
*Email: fyen@hit.edu.cn, renpeng@hit.edu.cn

**Supplementary Information:**

1. Experimental details
2. Tables listing the doping concentrations δ and molar magnetic susceptibilities $\chi_{Mol}$ of Examples 1-8 (**Tables S1-S8**).
3. List of experimentally obtained $\chi_{Mol}$ (**Table S9**).
4. Determination of δ of $NH_{4(1-\delta)}D_\delta Br$ (Example 1) and $NH_4I_{1-\delta}Br_\delta$ (Example 2) from lattice parameters (**Figures S1** & **S2**).
5. Determination of δ of $(NH_4H_2)_{1-\delta}(ND_4D_2)_\delta PO_4$ (Example 3) via measurements of the critical temperatures of the order-disorder phase transition (**Figure S3** & **S4**).

## 1. Experimental details:

### a) Magnetic susceptibility measurements:

The magnetic susceptibility was measured by the vibrating sample magnetometer (VSM) option of a Physical Properties Measurement System (PPMS) manufactured by Quantum Design, Inc. All measurements were performed at 300.00 K. The ramping speed of the applied magnetic field was set to 20 Oe/s. GE varnish was used to secure the samples onto the sample rods. The mass of the varnish was always measured so its diamagnetic contribution can be subtracted.

### b) Sample preparation:

$NH_{4(1-\delta)}D_{4\delta}Br$: Ammonium bromide $NH_4Br$ (CAS#: 12124-97-9, 99.99% in purity) was mixed in different ratios of deionized $H_2O$ and $D_2O$ (99.5% deuterium) to form a solution. After slow evaporation, the formed crystals were collected, grounded to powder and compressed into pellets of sizes 5 mm in diameter and 2 mm in height.

$NH_4I_{1-\delta}Br_\delta$: $NH_4Br$ and ammonium iodide $NH_4I$ (CAS#: 12027-06-4, 99.999%) were mixed in deionized $H_2O$ to form a solution. The slow evaporation method was also used and the mixed crystals were grounded and compressed into pellets.

$(NH_4H_2)_{1-\delta}(ND_4D_2)_\delta PO_4$: Ammonium dihydrogen phosphate $(NH_4)H_2PO_4$ (CAS#: 7722-76-1, ≥99.99%) was dissolved in different ratios of $H_2O$ and $D_2O$ (starting solutions are listed in Table S3). After slow evaporation, the collected crystals were grounded and pressed into pellet-form.

Pure solids of $NH_4I$, $NH_4Br$ and $NH_4H_2PO_4$: powder was taken straight from the reagent bottles and compressed into disc-shaped pellets.

For the remaining five series of examples, different quantities of the ten parent compounds were dry mixed, finely grounded into powder and pressed into pellets. Their control δ were determined from the masses of the parent compounds prior to mixing. The parent compounds were: Zirconium 1,4-dicarboxybenzene MOF, $C_{48}H_{28}O_{32}Zr_6$ (UiO-66, CAS#: 1072413-89-8, ≥97% in purity); $C_{48}H_{22}Br_6O_{32}Zr_6$ (UiO-66-Br, CAS#: 1260119-02-5, 97%); creatine, $C_4H_9N_3O_2$ (CAS#: 57-00-1, 98%); D-glucose, $C_6H_{12}O_6$ (CAS#: 50-99-7, ≥99.5%), L-glutamic acid, $C_5H_9NO_4$ (CAS#: 56-86-0, ≥99.5%); L-leucine, $C_6H_{13}NO_2$ (CAS#: 61-90-5, 99%); terephthalic acid, $C_8H_6O_4$ (CAS#: 100-21-0, 99%); trimesic acid, $C_9H_6O_6$ (CAS#: 554-95-0, 98%); *p*-terphenyl, $C_{18}H_{14}$ (CAS#: 92-94-4, ≥99.5%); and triphenylphosphine, $C_{18}H_{15}P$ (CAS#: 603-35-0, >99%).

### c) Lattice constant measurements:

The lattice constants were obtained by using a Rigaku XtaLAB CCD diffractometer equipped with graphite-monochromated Mo Kα radiation (λ = 0.71073 Å) via the ω-φ scanning technique. Single crystals taken directly from the mother liquor were placed in paraffin oil for their analyses. The crystals were mounted on a nylon loop and placed on the goniometer for data collection. All measurements were performed at 300 K.

## 2. Tables listing the doping concentrations δ and molar magnetic susceptibilities $\chi_{Mol}$ of examples 1-8.

The obtained molar magnetic susceptibility $\chi_{Mol}$ and doping concentration δ of the series of samples in Examples 1 to 8 (Figures 4 to 11) are listed in Tables S1 to S8, respectively. $\chi_{Mass}$ is the measured mass magnetic susceptibility for every sample. A comparison of δ obtained with the magnetic susceptibility method and control is also provided in the last columns. The difference in δ obtained from the current method and from a second control method was on average 1.86%, which is nearly twice as the theoretical expected value. This discrepancy may be attributed to uncertainties of the control method and errors introduced during the weighing process of the samples.

| δ in starting solution | Sample mass (mg) | $\chi_{Mass}$ (emu/g-Oe ×$10^{-6}$ ) | $\chi_{Mol}$ (cm$^3$/mol ×$10^{-6}$) | δ from $\chi_{Mol}$ | δ from lattice constant | δ-$\chi_{Mol}$ / δ-control % |
|---|---|---|---|---|---|---|
| 0.00 | 35.4 | – 0.50066(6) | – 49.036(6) | 0 | 0 | -- |
| 0.25 | 17.2 | – 0.50200(12) | – 49.555(10) | 0.192(5) | 0.187 | 2.67 |
| 0.67 | 16.4 | – 0.50495(12) | – 50.744(61) | 0.634(25) | 0.626 | 1.28 |
| 0.83 | 39.8 | – 0.50564(5) | – 50.972(55) | 0.719(22) | 0.747 | – 3.75 |
| 0.999×3 | 14.7 | – 0.50718(14) | – 51.674(14) | 0.980 | 0.980 | -- |

**Table S1.** Measured values of the $NH_{4(1-\delta)}D_{4\delta}Br$ series: mass magnetic susceptibility $\chi_{Mass}$, molar magnetic susceptibility $\chi_{Mol}$ and doping concentration δ from two methods. "δ in starting solution" takes into account the molar quantities of the reagent: $H_2O$ plus $D_2O$. The last column contrasts the obtained values of δ via the $\chi_{Mol}$ and XRD methods by taking their ratio minus one and multiplying by 100.

| δ in starting solution | Sample mass (mg) | $\chi_{Mass}$ (emu/g-Oe ×$10^{-6}$ ) | $\chi_{Mol}$ (cm$^3$/mol ×$10^{-6}$) | δ from $\chi_{Mol}$ | *a* from XRD (Å) |
|---|---|---|---|---|---|
| 0.00 | 41.6 | – 0.44133(5) | – 63.968(7) | 0 | 7.2758 |
| 0.06 | 12.8 | – 0.44381(16) | – 63.065(87) | 0.060(6) | 7.2530 |
| 0.09 | 18.3 | – 0.44435(11) | – 62.870(67) | 0.073(4) | 7.2481 |
| 0.25 | 16.4 | – 0.45087(12) | – 60.671(73) | 0.221(4) | 7.1859 |
| 0.98 | 25.6 | – 0.49588(8) | – 49.871(68) | 0.944(3) | 4.0590 |
| 1.00 | 35.4 | – 0.50066(6) | – 49.036(6) | 1 | 4.0659* |

**Table S2.** Measured values of the $NH_4I_{1-\delta}Br_\delta$ series; *value from Ref. 28.

| δ in starting solution | Sample mass (mg) | $\chi_{Mass}$ (emu/g-Oe ×$10^{-6}$ ) | $\chi_{Mol}$ (cm$^3$/mol ×$10^{-6}$) | δ from $\chi_{Mol}$ | δ from $T_C$ | δ-$\chi_{Mol}$ / δ-control % |
|---|---|---|---|---|---|---|
| 0.00 | 13.1 | – 0.50177(15) | – 57.717(18) | 0 | 0 | -- |
| 0.20 | 21.4 | – 0.50695(9) | – 58.692(34) | 0.124(7) | 0.120 | 3.33 |
| 0.40 | 6.5 | – 0.51296(31) | – 59.853(60) | 0.274(10) | 0.261 | 4.98 |
| 0.60 | 8.9 | – 0.51578(22) | – 60.448(73) | 0.351(8) | 0.336 | 4.46 |
| 0.80 | 7.8 | – 0.52678(26) | – 62.640(99) | 0.644(11) | 0.644 | 0.00 |

**Table S3.** Measured values of the $(NH_4H_2)_{1-\delta}(ND_4D_2)_\delta PO_4$ series.

| δ in starting mixture | Sample mass (mg) | $\chi_{Mass}$ (emu/g-Oe ×$10^{-6}$ ) | $\chi_{Mol}$ (cm$^3$/mol ×$10^{-6}$) | δ from $\chi_{Mol}$ | δ-$\chi_{Mol}$ / δ-control % |
|---|---|---|---|---|---|
| 0.000 | 6.5 | – 0.35796(31) | – 765.11(66) | 0 | -- |
| 0.497 | 9.9 | – 0.40877(20) | – 776.83(107) | 0.501(7) | 0.80 |
| 0.720 | 11.3 | – 0.43495(18) | – 781.90(125) | 0.718(7) | – 0.28 |
| 0.852 | 10.4 | – 0.45481(19) | – 788.70(141) | 0.867(8) | 1.76 |
| 1.000 | 10 | – 0.47386(20) | – 788.52(33) | 1.000 | -- |

**Table S4.** Measured values of the $[\text{UiO-66-Br}]_{(1-\delta)}[\text{UiO-66}]_\delta$ series.

| δ in starting mixture | Sample mass (mg) | $\chi_{Mass}$ (emu/g-Oe ×$10^{-6}$ ) | $\chi_{Mol}$ (cm$^3$/mol ×$10^{-6}$) | δ from $\chi_{Mol}$ | δ-$\chi_{Mol}$ / δ-control % |
|---|---|---|---|---|---|
| 0.000 | 14.3 | – 0.57157(14) | – 74.950(18) | 0 | -- |
| 0.223 | 11.0 | – 0.57770(18) | – 82.892(426) | 0.217(14) | – 2.69 |
| 0.410 | 17.2 | – 0.58260(12) | – 88.311(466) | 0.417(16) | 1.71 |
| 0.880 | 20.0 | – 0.59230(10) | – 104.032(375) | 0.908(15) | 3.19 |
| 1.000 | 14.0 | – 0.59383(14) | – 106.98(3) | 1.000 | -- |

**Table S5.** Measured values of the $[\text{creatine}]_{(1-\delta)}[\text{D-glucose}]_\delta$ series.

| δ in starting mixture | Sample mass (mg) | $\chi_{Mass}$ (emu/g-Oe ×$10^{-6}$ ) | $\chi_{Mol}$ (cm$^3$/mol ×$10^{-6}$) | δ from $\chi_{Mol}$ | δ-$\chi_{Mol}$ / δ-control % |
|---|---|---|---|---|---|
| 0.000 | 14.0 | – 0.55413(14) | – 81.529(21) | 0 | -- |
| 0.146 | 13.6 | – 0.57626(15) | – 83.430(24) | 0.147(2) | 0.68 |
| 0.590 | 9.6 | – 0.64620(21) | – 89.057(36) | 0.581(3) | – 1.52 |
| 0.651 | 13.3 | – 0.65777(15) | – 89.936(33) | 0.649(3) | – 0.31 |
| 1.000 | 13.3 | – 0.72033(15) | – 94.486(20) | 1.000 | -- |

**Table S6.** Measured values of the $[\text{L-glutamic acid}]_{(1-\delta)}[\text{L-leucine}]_\delta$ series.

| δ in starting mixture | Sample mass (mg) | $\chi_{Mass}$ (emu/g-Oe ×$10^{-6}$ ) | $\chi_{Mol}$ (cm$^3$/mol ×$10^{-6}$) | δ from $\chi_{Mol}$ | δ-$\chi_{Mol}$ / δ-control % |
|---|---|---|---|---|---|
| 0.000 | 23.0 | – 0.52028(9) | – 86.434(14) | 0 | -- |
| 0.170 | 17.0 | – 0.50826(12) | – 88.276(62) | 0.172(4) | 1.18 |
| 0.632 | 19.4 | – 0.48052(10) | – 93.243(109) | 0.634(6) | 0.32 |
| 0.685 | 13.9 | – 0.47819(14) | – 93.866(122) | 0.678(7) | – 1.02 |
| 1.000 | 12.5 | – 0.46240(16) | – 97.168(33) | 1.000 | -- |

**Table S7.** Measured values of the [terephthalic acid]$_{(1-\delta)}$[trimesic acid]$_{\delta}$ series.

| δ in starting mixture | Sample mass (mg) | $\chi_{Mass}$ (emu/g-Oe ×$10^{-6}$ ) | $\chi_{Mol}$ (cm$^3$/mol ×$10^{-6}$) | δ from $\chi_{Mol}$ | δ-$\chi_{Mol}$ / δ-control % |
|---|---|---|---|---|---|
| 0.000 | 13.7 | – 0.68065(15) | – 178.54(4) | 0 | -- |
| 0.209 | 17.4 | – 0.68291(11) | – 174.34(98) | 0.219(48) | 4.78 |
| 0.281 | 14.8 | – 0.68359(14) | – 173.16(75) | 0.282(29) | – 0.36 |
| 0.589 | 17.1 | – 0.68676(12) | – 167.68(85) | 0.567(40) | – 3.74 |
| 1.000 | 11.8 | – 0.69208(17) | – 159.39(4) | 1.000 | -- |

**Table S8.** Measured values of the [triphenylphosphine]$_{(1-\delta)}$[*p*-terphenyl]$_{\delta}$ series.

## 3. List of experimentally obtained $\chi_{Mol}$.

| Compound name | $\chi_{Mol}$ (cm$^3$/mol ×$10^{-6}$ ) (Experimental) | $\chi_{Mol}$ (cm$^3$/mol ×$10^{-6}$) (Expected) | Sample purity (%) |
|---|---|---|---|
| $NH_4Br$ | – 49.036(6) | – 47.9 | 99.99 |
| $ND_4Br$ | – 51.674(14)* | N/A† | 100 -*d* |
| $NH_4I$ | – 63.968(7) | – 63.9 | 99.999 |
| $NH_4H_2PO_4$ | – 57.717(18) | – 63.86 | 99.99 |
| $ND_4D_2PO_4$ | – 65.402(30)* | N/A† | 100 -*d* |
| $C_{48}H_{28}O_{32}Zr_6$ | – 788.52(33) | N/A‡ | ≥97 |
| $C_{48}H_{22}Br_6O_{32}Zr_6$ | – 765.11(66) | N/A‡ | >97 |
| Creatine | – 74.950(18) | – 77.39 | 98 |
| D-glucose | – 106.98(3) | – 101.5 | ≥99.5 |
| L-glutamic acid | – 81.529(21) | – 78.5 | ≥99.5 |
| L-leucine | – 94.486(20) | – 84.9 | 99 |
| terephthalic acid | – 86.434(14) | – 84.22 | 99 |
| Trimesic acid | – 97.168(33) | – 97.68 | 98 |
| Triphenylphosphine | – 178.54(4) | – 187.0 | >99 |
| *p*-terphenyl | – 159.39(4) | – 157.8 | ≥99.5 |

**Table S9. List of $\chi_{Mol}$ values of the parent compounds obtained experimentally at 300 K.** The expected values of $\chi_{Mol}$ were calculated from addition of Pascal's constants taken from Ref. 26.

*98% deuteration.

†Pascal's constants of $ND_4$ not available in existing literature.

‡Pascal's constant of the –Zr covalent bond also not available.

## 4. Determination of δ of the $NH_{4(1-\delta)}D_{\delta}Br$ series (Example 1) and $NH_4I_{1-\delta}Br_{\delta}$ series (Example 2) from lattice parameters.

The doping concentration δ was also extracted from measured lattice parameters of $NH_{4(1-\delta)}D_{\delta}Br$ (Figure S1) and $NH_4I_{1-\delta}Br_{\delta}$ (Figure S2) as a means to check the accuracy of the obtained δ from our magnetic measurements.

The lattice constants were obtained by using a Rigaku XtaLAB CCD

diffractometer equipped with graphite-monochromated Mo Kα radiation (λ = 0.71073 Å) via the ω-φ scanning technique. Single crystals taken directly from the mother liquor were placed in paraffin oil for their analyses. The crystals were mounted on a nylon loop and placed on the goniometer for data collection. All measurements were performed at 300 K.

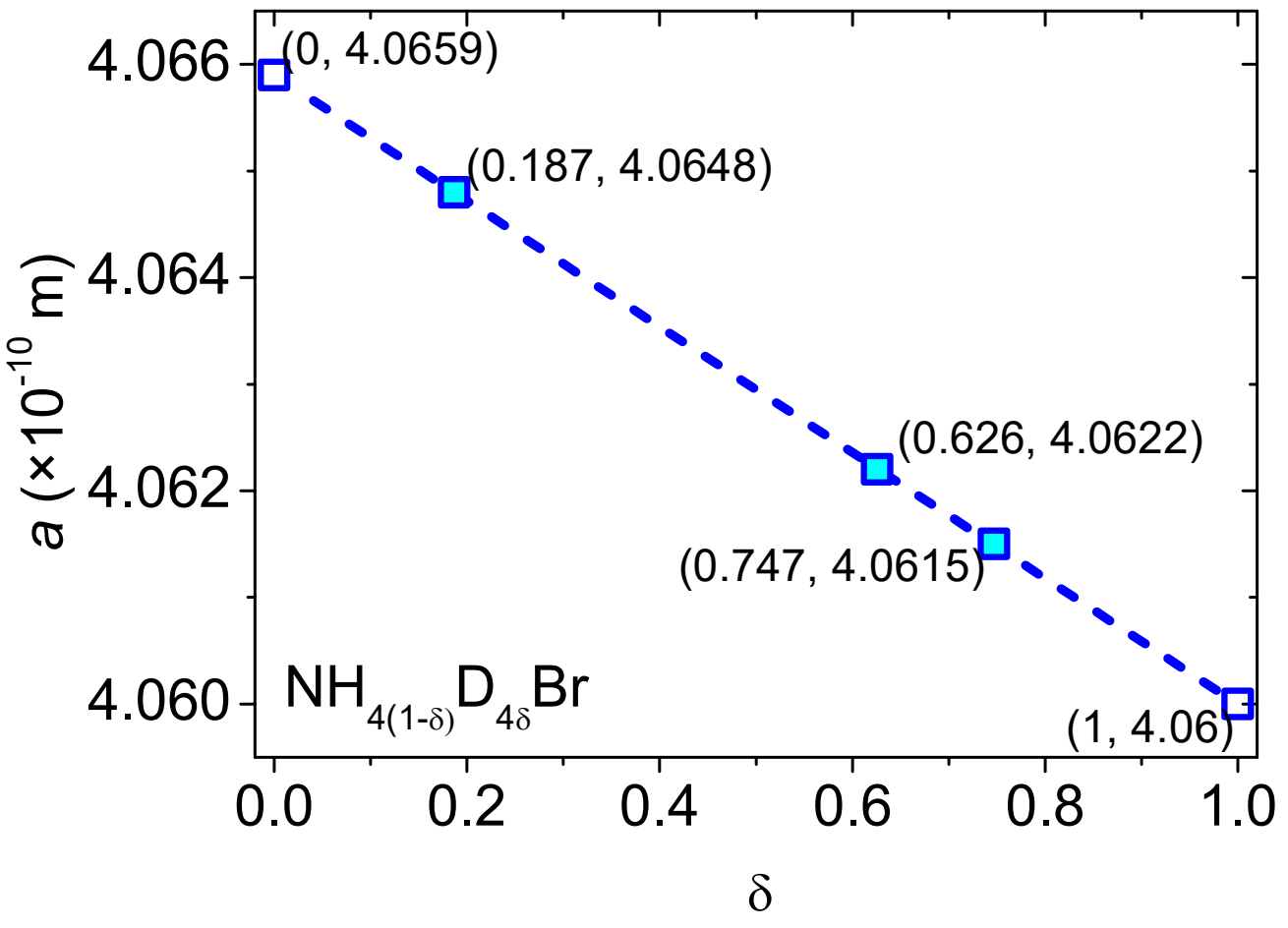

**Fig. S1.** Lattice constant *a* vs. isotope concentration. Open squares represent the lattice constants of $NH_4Br$ and $ND_4Br$ according to Levy *et al.*[28] and Havighurst *et al.*[9] The three solid squares are the obtained lattice constants of the three mixed solids and their corresponding δ.

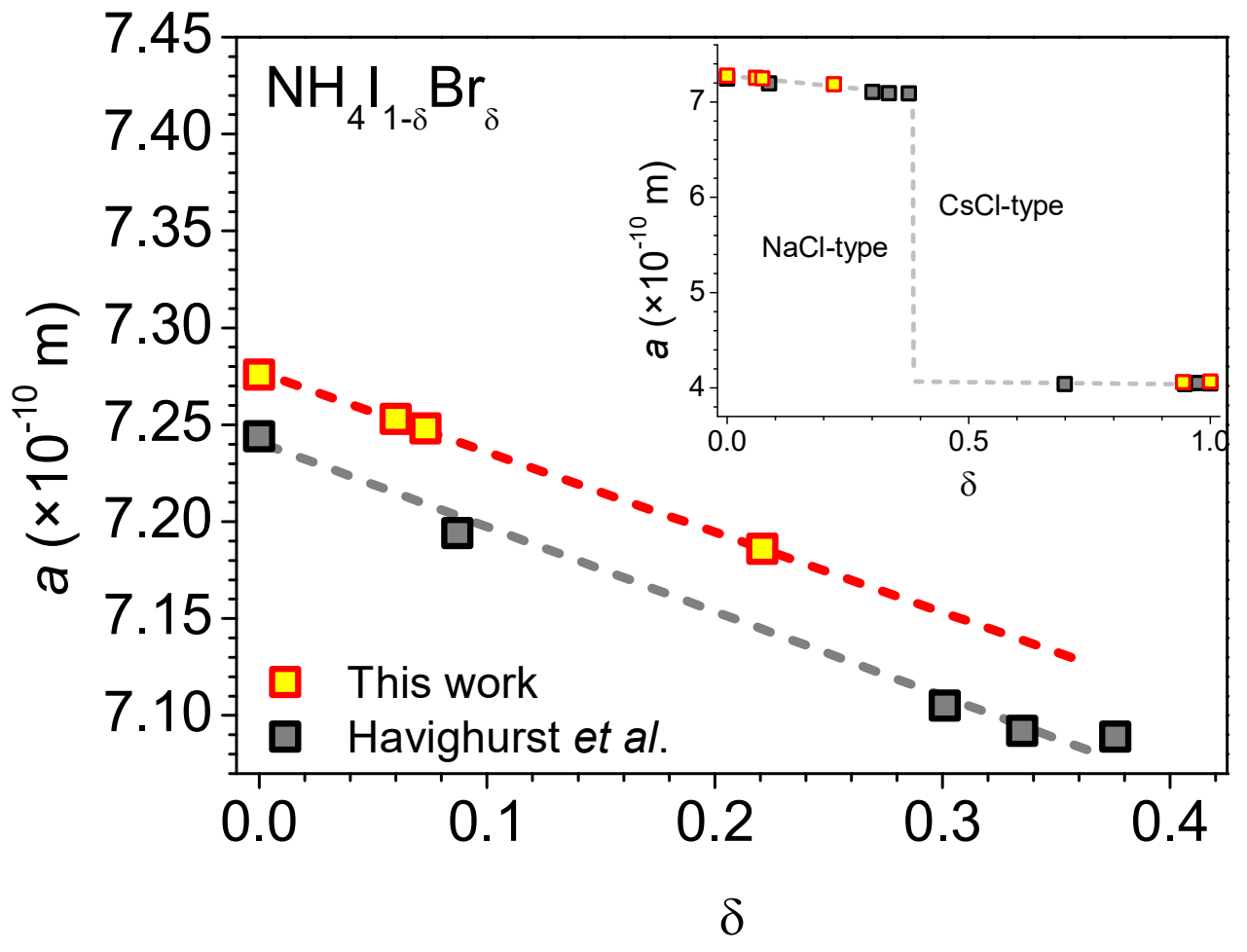

**Fig. S2.** Lattice constant *a* vs. halide concentration. Yellow squares are data from the last two columns of Table 2. Grey squares are data from Havighurst *et al.*[9] Dashed lines are linear fits. Inset shows how *a* is no longer additive when δ > 0.37.

## 5. Determination of δ of the $(NH_4H_2)_{1-\delta}(ND_4D_2)_\delta PO_4$ series (Example 3) via measurements of the critical temperatures of the order-disorder phase transition.

For the series $(NH_4H_2)_{1-\delta}(ND_4D_2)_\delta PO_4$, δ was extracted from the critical temperature $T_C$ of its order-disorder phase transition according to $\delta = (T_C - 148\text{ K}) / 94\text{ K}$.[33] Figure S3 shows the cooling and warming runs at 2 K/min of the mixed samples with their respective $T_C$'s. Figure S4 is a graphical representation of the linear relationship between the measured $T_C$'s with δ. In each case, δ was obtained from the average $T_C$ value of the cooling and warming curves.

The dielectric constants of the mixed crystals of the series $(NH_4H_2)_{1-\delta}(ND_4D_2)_\delta PO_4$ were obtained from their measured capacitances with an E4980A LCR meter from Agilent Technologies. The electrodes were in to form of silver paint applied onto the surfaces perpendicular to the *b*-axis direction of the crystals. The applied electric field and frequency were ~5 V/cm and 1 kHz, respectively.

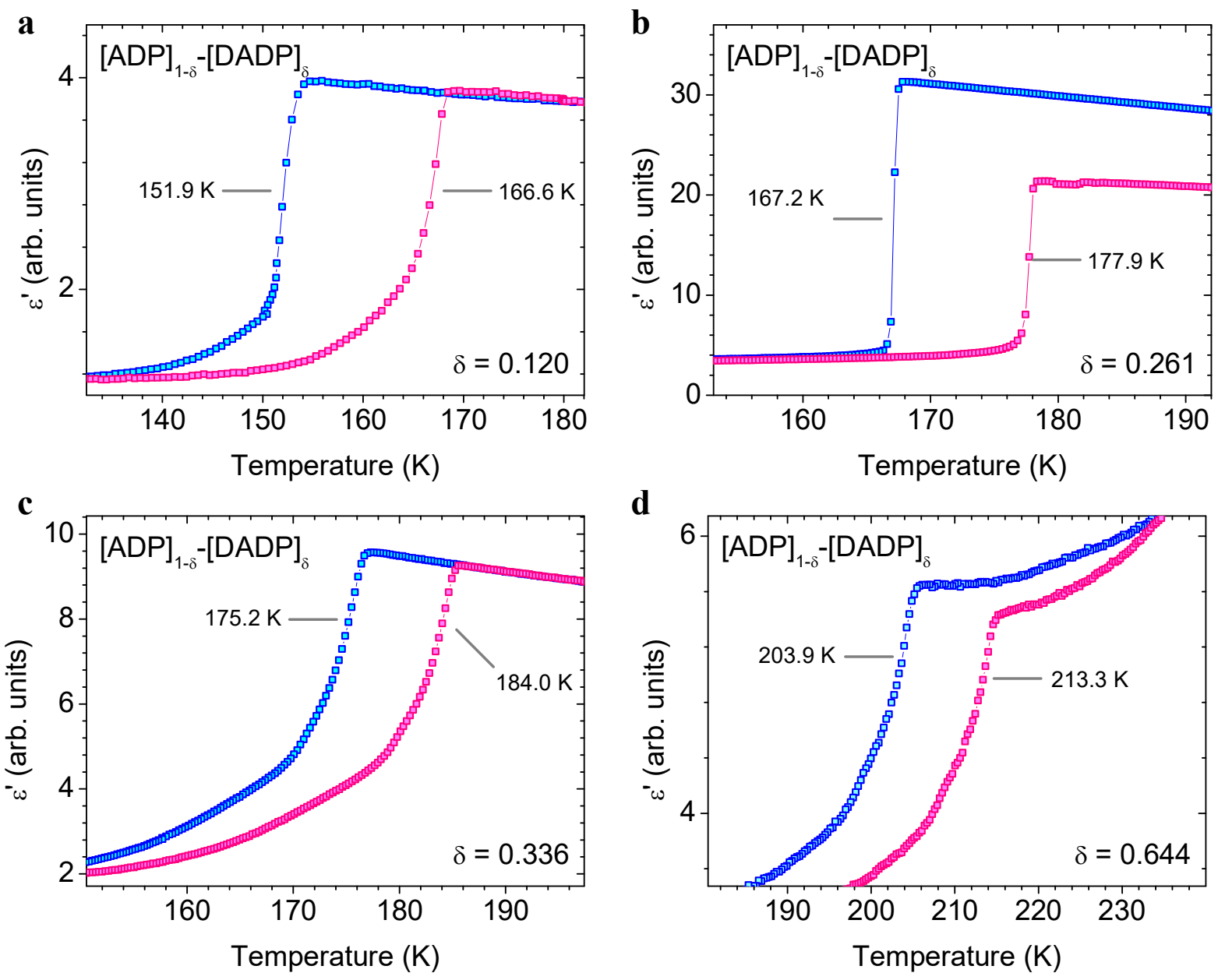

**Fig. S3.** Dielectric constant of the series $(NH_4H_2)_{1-\delta}(ND_4D_2)_\delta PO_4$, $[ADP]_{1-\delta}$-$[DADP]_\delta$, with respect to temperature. The obtained δ for the cases when a) δ = 0.120; b) δ = 0.261; c) δ = 0.336; and d) δ = 0.644 were obtained from the average of the critical temperatures $T_C$ during cooling and warming for each sample.

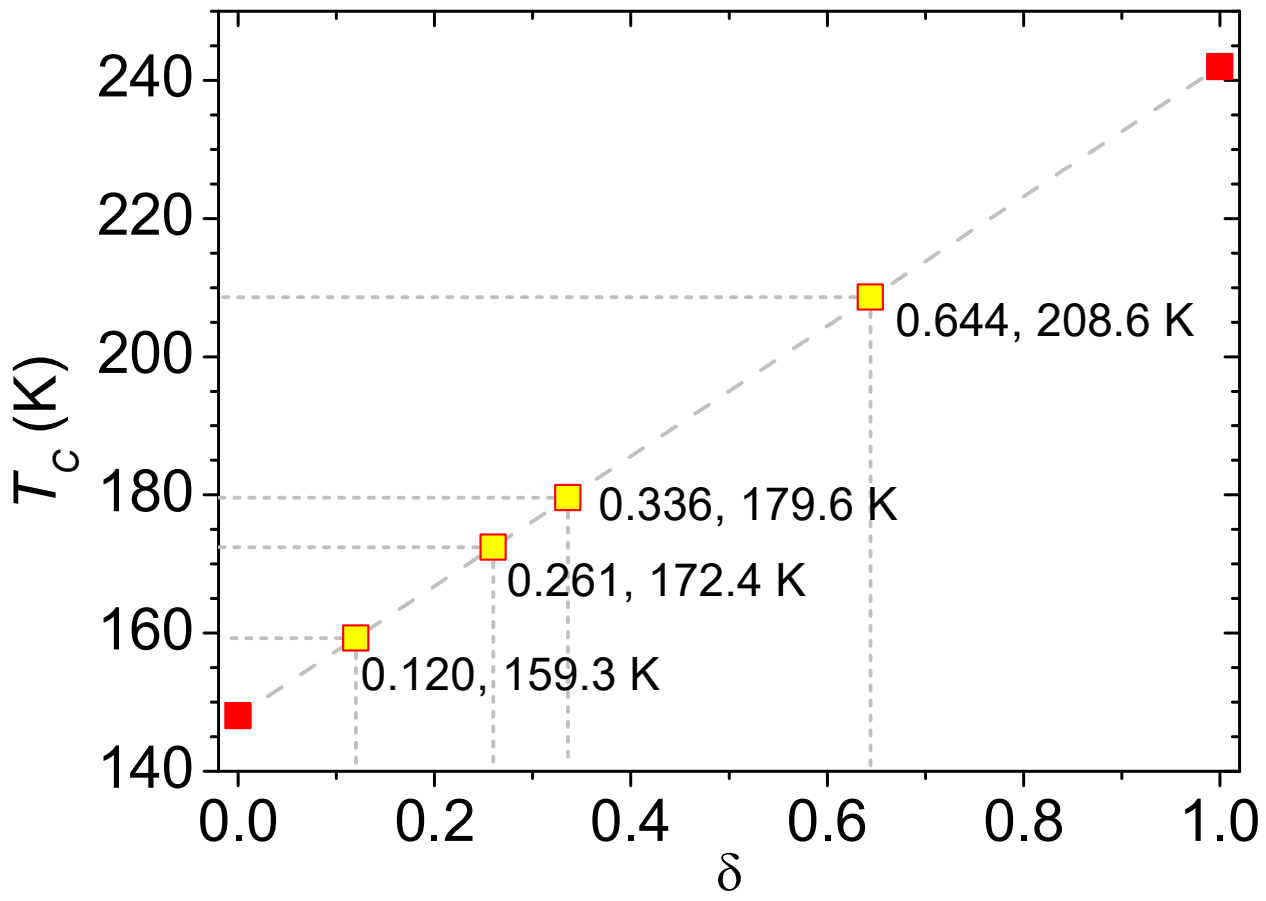

**Fig. S4.** The equation $\delta = (T_C - 148\text{ K}) / 94\text{ K}$ was employed to obtain δ from $T_C$. Red squares represent the parent compounds $NH_4H_2PO_4$ and $ND_4D_2PO_4$; yellow squares the four mixed samples.